\documentclass[11pt,letterpaper]{article}
\pdfoutput=1

\usepackage{jcappub}
\usepackage[config=altsf]{subfig}
\usepackage{feynmf}
\usepackage{array}
\usepackage{slashed}
\usepackage{afterpage}
\usepackage{appendix}
\usepackage{multirow}
\usepackage{mdframed}

\newenvironment{aside}
  {\begin{mdframed}[style=0,%
      leftline=false,rightline=false,leftmargin=2em,rightmargin=2em,%
          innerleftmargin=0pt,innerrightmargin=0pt,linewidth=0.75pt,%
      skipabove=7pt,skipbelow=7pt]\small}
  {\end{mdframed}}

\newcommand{\pa}{\partial}

\newcommand{\be}{\begin{equation}}
\newcommand{\ee}{\end{equation}}
\newcommand{\ba}{\begin{eqnarray}}
\newcommand{\ea}{\end{eqnarray}}

\newcommand{\en}{\nonumber\\}

\newcommand{\de}{\delta}

\newcommand{\kk}{\mathbf{k}}
\newcommand{\xx}{\mathbf{x}}

\newcommand{\p}{\mathbf{p}}

\newcommand{\pp}[1]{\mathbf{p}_{#1}}

\definecolor{darkred}{RGB}{175,0,0}
\definecolor{darkblue}{RGB}{0,0,175}
\newcommand{\fD}{f_{X}}

\usepackage{color}

\definecolor{dgreen}{rgb}{0,0.6,0.1}

\definecolor{purple}{rgb}{0.8, 0.2, 0.8}

\title{Dark Catalysis}
\author{Prateek Agrawal,}
\author{Francis-Yan Cyr-Racine,}
\author{Lisa Randall}
\author{and\\ Jakub Scholtz}
\affiliation{Department of Physics, Harvard University, \\
17 Oxford St., Cambridge, MA 02138, USA}

\emailAdd{prateekagrawal@fas.harvard.edu, fcyrraci@physics.harvard.edu,  randall@physics.harvard.edu, jscholtz@physics.harvard.edu}

\abstract{ Recently it was shown that dark matter with mass of order
  the weak scale can be charged under a new long-range force,
  decoupled from the Standard Model, with only weak constraints from
  early Universe cosmology. Here we consider the implications of an
  additional charged particle $C$ that is light enough to lead to
  significant dissipative dynamics on galactic times scales. We
  highlight several novel features of this model, which can be relevant even
when the $C$ particle constitutes only a small fraction of the number density
(and energy density). We assume a
  small asymmetric abundance of the $C$ particle whose charge is compensated by a heavy $X$ particle so that
 the relic abundance of dark matter consists mostly
  of symmetric $X$ and $\bar{X}$, with a small asymmetric component
  made up of $X$ and $C$. As the universe cools, it undergoes
  asymmetric recombination binding the free $C$s into $(XC)$ dark
  atoms efficiently. Even with a tiny asymmetric component, the
  presence of $C$ particles catalyzes tight coupling between the heavy
  dark matter $X$ and the dark photon plasma that can  lead to a
  significant suppression of the matter power spectrum on small scales
  and lead to some of the strongest bounds on such dark matter
  theories.  We find a viable parameter space where structure formation
  constraints are satisfied and significant dissipative dynamics can
  occur in galactic haloes but show a large region is excluded. 
  Our model shows that subdominant components in the dark sector can
  dramatically affect structure formation.
 
}
\arxivnumber{1702.05482}
\begin{document}
\maketitle

\section{Introduction}
Through its gravitational influence on luminous baryonic matter, dark
matter shapes the structures that we observe throughout the cosmos on
a broad range of length scales. While the cold dark matter paradigm
\cite{1981ApJ...250..423D,Blumenthal:1982mv,
Bond:1983hb,Blumenthal:1984bp,Davis:1985rj}
is in excellent agreement with observations on large cosmological
scales, the possibility that nongravitational dark matter physics
plays an important role on small astrophysical scales remains an
exciting possibility. Indeed, the presence of new interactions in the
dark matter sector can lead to observable differences in the dark
matter density distribution on small scales. They can also impact the
way dark matter is produced in the early Universe. Many possibilities
have already been explored in the literature
\cite{Goldberg:1986nk,HOLDOM198665,1992ApJ...398..407G,1992ApJ...398...43C,Foot:2003jt,Foot:2004pa,Foot:2004wz,Feng:2008mu,Ackerman:2008gi,Feng:2009mn,ArkaniHamed:2008qn,Kaplan:2009de,2010PhRvD..81h3522B,Kaplan:2011yj,Behbahani:2010xa,Das:2012aa,Hooper:2012cw,Aarssen:2012fx,Cline:2012is,Tulin:2013teo,Tulin:2012wi,Baldi:2012ua,Cyr-Racine:2013ab,Cline:2013zca,Baek:2013dwa,Chu:2014lja,Cline:2013pca,Bringmann:2013vra,Archidiacono:2014nda,Foot:2014mia,Foot:2014uba,Choquette:2015mca,Buen-Abad:2015ova,Ko:2016uft,2016PhRvD..93l3527C,Chacko:2016kgg,Ko:2016fcd,Kamada:2016qjo,Ko:2017uyb}.
One of the simplest possibilities is that dark matter is charged under
a new invisible $U(1)$ gauge group. Such a scenario is possible both
within the realm of symmetric \cite{Ackerman:2008gi,Feng:2009mn} and
asymmetric
\cite{Foot:2002iy,Kaplan:2009de,Kaplan:2011yj,Cyr-Racine:2013ab,Cline:2013zca}
dark matter theories, with the latter possibly leading to the
formation of composite dark atoms at late times.  

The presence of a massless dark photon in such scenarios can
dramatically affect the assembly and evolution of dark matter
structures on small scales. For instance, frequent scattering between
dark matter and dark photons has the potential to damp the growth of matter structure
\cite{Boehm:2000gq,Boehm:2001hm,Boehm:2004th,Feng:2009mn,Cyr-Racine:2013ab,Schewtschenko:2014fca,Buckley:2014hja,Foot:2016wvj}
on small scales, affecting the spectrum of cosmic microwave background
(CMB) anisotropies \cite{Foot:2011ve,Cyr-Racine:2013ab,Cyr-Racine:2013fsa} and
leading to a suppressed number of satellites orbiting typical galaxies
today. In addition, the presence of a light particle coupled to the
massless dark photon can provide a channel for dark matter to shed
energy and momentum \citep{Foot:2013vna,Foot:2014mia,Foot:2013lxa,Foot:2015sia,Foot:2014uba,Foot:2015mqa,Foot:2016wvj,Chashchina:2016wle,Foot:2017dgx,Agrawal:2017pnb,DAmico:2017lqj} allowing for the formation of dense dark matter
objects such as disks \cite{Fan:2013yva,Fan:2013tia}. Moreover, the
dark matter self-interactions that are inherent to these
$U(1)$-charged models can lead to halo evaporation
\cite{Kahlhoefer:2013dca}. Also, self-interaction can lead to cored
dark matter halos \cite{2016MNRAS.460.1399V} which, being less dense,
are more susceptible to tidal disruption \cite{2016MNRAS.461..710D}. The above considerations do not in general constrain models with particles masses above several hundred GeV, leaving a fairly large
swath of parameter space where these theories are viable
\cite{Agrawal:2016quu}. 

The above constraints could also be satisfied, even in the presence of lighter dark matter, if only a
fraction of the dark matter is charged under the new $U(1)$ force, as
was considered in refs.~\cite{Fan:2013yva,Fan:2013tia,Cyr-Racine:2013fsa,Chacko:2016kgg}.
These latter models are appealing since they can retain the success of
the standard cold dark matter paradigm while introducing genuinely new
phenomenology on small scales. However, for models with eV-scale binding energy
between the darkly charged constituents, constraints from CMB
measurements limit the fraction of this charged matter to about $5\%$
\cite{Cyr-Racine:2013fsa} when decoupling of the two sectors happens
at the weak scale and the temperature of the dark sector is about half
that of the Standard Model sector. This latter constraint is somewhat
weaker for a cooler dark sector or for models with higher binding
energy since the formation of neutral dark atoms happens earlier in
these cases. 

In this paper, we consider the perhaps more economical scenario than
that proposed in refs.~\cite{Fan:2013yva,Fan:2013tia}, in which all
the dark matter is charged under an invisible new $U(1)$ force, while
retaining the interesting small-scale phenomenology explored in these
papers, by allowing for the presence of a light particle charged under the
new dark force. In detail, we explore here an asymmetry between the darkly-charged particles $X$ and
$\bar{X}$ forming the bulk of the dark matter density that is exactly
compensated by the presence of the light darkly-charged particles $C$. This model has interesting phenomenological implications independent of the double-disk dark matter (DDDM) scenario \cite{Fan:2013yva,Fan:2013tia}. Within this latter model, the economical scenario considered
here potentially allows the interesting possibility that the dark matter within
the galactic halo and the dark disk have related composition.  We show that such a light $C$ particle can have a tremendous effect on dark matter structure formation despite
contributing negligibly to the overall mass density of dark matter.
Compared to the original DDDM scenario, structure formation
constraints on this new model are generally more stringent, even for
relatively heavy dark matter, because the dark matter sector can be
entirely coupled to the dark radiation bath through the light
darkly-charged particle. We show that there is
nevertheless a valid parameter region where the model is viable. 

The outline of this paper is as follows. In section \ref{sec:model},
we describe the main ingredients entering our dark matter models and
highlight its most relevant implications. In section
\ref{sec:relic-abundance}, we present the dark matter relic abundance
calculation in the presence of the asymmetry. In section
\ref{sec:recombination}, we compute the cosmological evolution of the
ionized fraction of the light darkly-charged particle in the presence
of the asymmetric bath of heavy $X$ and $\bar{X}$ particles. We make
use of this latter result in section \ref{sec:structure_formation} to
study the evolution and growth of dark matter fluctuations within this
model. In section~\ref{sec:Hubble} we study the $H_0$ determination
from the CMB in presence of the dark photons
appearing as
additional relativistic degrees of freedom.
In section \ref{sec:late-time-astro}, we discuss the late-time
astrophysical consequences of the model. We finally conclude in section \ref{sec:conclusions}.

\section{Model and Implications}\label{sec:model} 
The model consists simply of a heavy charged particle $X$ carrying
positive charge under a new dark $U(1)$ gauge group, its antiparticle
$\bar{X}$ with opposite charge, and a light particle $C$ that also
carries negative charge. For concreteness we will consider $X$ and $C$
to be Dirac fermions throughout this paper. The Lagrangian is
\be
\mathcal{L}  =  -\frac14 V_{\mu\nu} V^{\mu\nu} + \bar{X} \slashed{D} X +\bar{C}\slashed{D}C - m_X \bar{X} {X} - m_C\bar{C}C,
\ee
where $V_{\mu\nu}$ is the dark photon field strength, $D$ is the gauge
covariant derivative, and $m_X$ and $m_C$ are the masses of the $X$
and $C$ particles, respectively. We denote the fine-structure constant
of the new $U(1)$ force as $\alpha_D$. In general, the temperature $T_D$ of the dark photon bath will be different than that of the Standard Model sector, and we denote by $\xi(T) \equiv T_D/T$ the ratio of the dark sector to the Standard Model temperature (denoted by $T$ throughout our paper). The evolution of the dark photon temperature is discussed further in appendix \ref{app:temp_evol}. In our scenario, $X$ and
$\bar{X}$ act as the usual cold dark matter that forms the
gravitational backbone of galaxies, clusters, and the large-scale
structure of the Universe, whereas the light $C$ particle can act as a
catalyst between the bulk of the dark matter density and the new dark
photons, allowing the damping of small-scale dark matter fluctuations
and the cooling of dark matter particles within galactic halos. We
emphasize that this differs from the model of
refs.~\cite{Fan:2013tia,Fan:2013yva} in that the bulk of the dark
matter is now charged under the new $U(1)$ force. We summarize here
the key phenomenological implications of this new model.  
\subsection{Distinctive features of the model }
\paragraph{Relic abundance}
As we will describe in detail in section \ref{sec:relic-abundance},
the relic abundance calculation has notable differences from the
standard thermal freeze-out in the presence of a long-range force (see
e.g.~refs.~\cite{Ackerman:2008gi,Feng:2009mn}). In particular, we
identify two distinct regimes of interest. For small values
of $\alpha_D$ and of the asymmetry,  the relic abundance is set
through standard thermal freeze-out with the asymmetry parameter
playing essentially no role, except for establishing the subdominant
abundance of the light $C$ particles. For large values of the
dark fine-structure constant, the overall dark matter relic abundance
is essentially entirely determined by the value of the asymmetry
parameter, and the calculation closely follows that of standard
asymmetric dark matter models \cite{Kaplan:2009ag}.  We will be particularly
interested in the transition region between these two regimes where
both the asymmetric and symmetric components are present with a
sizable abundance.   

\paragraph{Asymmetric recombination}
Once the relic abundance is established in the early Universe, the
$X$, $\bar{X}$ and $C$ particles form an ionized plasma interacting
with the dark photon bath. Once the dark sector temperature falls significantly below the binding
energy between the heavy $X$ and the light $C$, it becomes
energetically favorable for the plasma to form neutral $(XC)$ bound
states. The presence of the asymmetry between the total number of $X$ and $C$ particles in the thermal bath leads to a qualitatively different recombination process compared to that of standard hydrogen \cite{Peebles:1968ja,1968ZhETF..55..278Z}. We present a detailed
computation of this asymmetric recombination in section
\ref{sec:recombination}. We find that the abundance of free $C$
particles (that is, those not in bound states) strongly depends on the
value of the dark fine-structure constant, with values of $\alpha_D
\gtrsim 0.005$ generally resulting in a negligible population of free
$C$s at late times, while smaller values of $\alpha_D$ result in
little $(XC)$ recombination. The final free $C$ fraction at late times
also displays weak dependence on the masses of the $X$ and $C$
particles. 

\paragraph{Structure formation}
Similar to the case of the standard photon-baryon plasma, the presence of the light $C$ particles allows the coupling of dark matter to the dark photons, prohibiting the growth of dark
matter fluctuations on scales that enter the causal horizon before
dark matter kinetic decoupling
\cite{Boehm:2000gq,Boehm:2001hm,Boehm:2004th,Feng:2009mn,Cyr-Racine:2013ab,Cyr-Racine:2013fsa,Schewtschenko:2014fca,Buckley:2014hja}. This tight coupling arises from 
frequent Coulomb-type interactions between
the darkly-charged constituents which maintain kinetic equilibrium among
these particles, while Thomson scattering between dark photons and
light $C$ particles is responsible for establishing kinetic
equilibrium between the dark matter and the dark radiation.

An important constraint of the model comes from demanding that dark
matter decouples early enough from the dark photon bath so as not to
erase density fluctuations on scales where we have direct
observations. The observed abundance of satellite galaxies in the
Local Group \cite{2010MNRAS.406.1220W} and measurements of the
Lyman-$\alpha$ forest absorption spectrum
\cite{2006PhRvL..97s1303S,2013PhRvD..88d3502V,2015JCAP...11..011P,2016JCAP...08..012B}
put the strongest constraints on the epoch of dark matter kinetic
decoupling. While these constraints probe the nonlinear regime of
structure formation and are therefore subject to modeling
uncertainties, both can be roughly captured by demanding that the
matter power spectrum does not significantly deviate from its standard
cold dark matter behavior on comoving scales corresponding to $k < 10
h/$Mpc. 

Other key observable signatures of this scenario include the presence of dark acoustic oscillations \citep{1992ApJ...398..407G,Cyr-Racine:2013ab,Cyr-Racine:2013fsa} in the dark matter fluid on scales entering the horizon before the epoch of dark matter kinetic decoupling. Depending on the exact choice of parameters, these dark oscillations could leave imprints on the subhalo mass function (see e.g.~ref.~\cite{Buckley:2014hja}) that are distinct from those of warm dark matter. Structure formation
constraints are discussed further in section
\ref{sec:structure_formation} below.

\paragraph{Dark photons and Hubble rate}
When comparing our model to current CMB data \cite{Ade:2015xua,2016JCAP...01..007B}, the presence of the dark radiation component allows for a larger value of the present-day Hubble parameter $H_0$ than that  obtained using a standard 6-parameter $\Lambda$CDM cosmology.  Our model can thus help to alleviate the tension between local $H_0$ measurements \cite{2016ApJ...826...56R} and those inferred from CMB and large-scale structure measurements. In addition, the CMB bound
 on the effective number of
relativistic species ($N_{\rm eff}$) can be translated to a
constraint on the temperature of the dark photon bath, generally
requiring the dark sector temperature to be at most half that of the
visible sector \cite{Cyr-Racine:2013fsa}. We elaborate on these constraints in section \ref{sec:Hubble}.

\paragraph{Late-time astrophysics: dissipative dynamics}
 For massive dark matter halos with a final virial temperature higher
than the $(XC)$ binding energy, shock heating will generally ionize the
$(XC)$ bound states, resulting in a net fraction of free $C$ particles
orbiting within dark matter halos. As was discussed in
refs.~\cite{Fan:2013tia,Fan:2013yva}, the presence of light
darkly-charged particles enable dissipative processes, such as
Bremsstrahlung and Thomson cooling,   which can alter the internal
structure of dark matter halos, or even lead to the formation of a
dark disk. In section \ref{sec:late-time-astro}, we reevaluate the
available parameter space where such phenomenology arises, and find
that it can lead to tensions with structure formation constraints. On
the one hand, efficient cooling within galactic halos requires the $C$
particles to be fairly light ($\lesssim 1$ MeV). On the other hand,
escaping the structure formation constraints is easiest when the $C$
particles are fairly heavy since in this case, dark recombination (and
kinetic decoupling) happens early enough as to not modify the matter
power spectrum on scales $k\lesssim 10h/$Mpc. This tension is
alleviated for large $X$ mass. Our analysis shows that it is possible
for dissipative dark matter dynamic to be compatible with structure
formation constraints for $m_C\sim 1$ MeV, $m_X\sim10$ TeV, and
$\alpha_D\sim0.15$. However, Bremsstrahlung and Thomson cooling do not
lead to disk formation in this model.

The dissipative dynamics present in the $(XC)$ model could also allow the formation of dense dark matter clouds \cite{Agrawal:2017pnb} where dark matter annihilation would be enhanced, hence potentially leading to indirect detection signatures within gamma-ray telescopes if a portal between the SM and the dark sector exists. Dissipation within the dark sector could also lead to the formation of black holes \cite{DAmico:2017lqj}.


\paragraph{Dark matter self-interactions}

The presence of long-range interactions lead to a strongly velocity-dependent self-interaction cross section among dark matter constituents within halos \cite{Agrawal:2016quu}. The resulting very large cross section within small-mass halos could lead to interesting phenomenology \cite{Ahn:2004xt,Moore:2000fp}. The strength of the dark matter self-interaction is constrained by observations of the detailed inner structure of halos and by studying the merger
of galaxy clusters. Constraints from the Bullet
cluster \cite{Randall:2007ph} and from subhalo evaporation
\cite{Kahlhoefer:2013dca} can be easily evaded for sufficiently heavy
$X$ particles \cite{Feng:2009mn}. Technically, bounds from the
triaxial structure of dark matter halos could offer even stronger
constraints on the dark matter self-interaction cross section.
However, as was discussed in ref.~\cite{Agrawal:2016quu}, a
combination of inaccurate modeling and poorly measured ellipticities
of dark matter halos effectively leads to a very weak bound on the
self-interaction cross section.

\section{Relic abundance}\label{sec:relic-abundance}

Here we derive the relic abundance of $X,\bar{X}$ and $C$ particles,
denoted collectively as $\Omega_X$.  The calculation of the relic
abundance proceeds in the standard way with two important
modifications. First, if the dark sector is kinetically decoupled from
the SM, we have to allow for a potential relative temperature in the
two sectors, $\xi(T)=T_D/T\neq1$. Secondly, we have to take into
account the asymmetry in $X$. Throughout this section, we generally
follow the formalism presented in refs.~\cite{Feng:2008mu} and
\cite{Graesser:2011wi}, but modify it appropriately to take into
account the above-mentioned temperature difference and asymmetry. 

We are particularly interested in models for which both the symmetric
and asymmetric components survive at late times. We note that
demanding the presence of the light $C$ requires a nonvanishing
asymmetry since its symmetric component would annihilate away
efficiently. We focus here on models where the asymmetric component is
a small fraction of the total dark matter energy density such that
the bulk of the dark matter is made up of the symmetric component $X$
and $\bar{X}$. If the interactions of $X$ with the SM are small, the
dominant annihilation channels for $X$ are in the dark sector, $X
\bar{X} \to \gamma_D \gamma_D$ and $X \bar{X} \to C \bar{C}$.

The Boltzmann equation for the evolution of the $X$ and $\bar{X}$
number density (denoted by $n_X$ and $n_{\bar{X}}$, respectively) can
be written as 
\begin{align} \frac{d n_{X,\bar{X}}}{dt} + 3 H
  n_{X,\bar{X}} &= -\langle \sigma v \rangle \left( n_X n_{\bar{X}} -
  n^{\rm eq}_X n^{\rm eq}_{\bar{X}} \right), 
\end{align}
where $H$ is the Hubble rate, $\langle \sigma v \rangle$ is the
thermally-averaged cross section times velocity (see below for
definition), and $n^{\rm eq}_X$ is the equilibrium value of the $X$
number density. As usual, we define the yield as
\begin{align}
  Y_{X,\bar{X}}
  &=
  \frac{n_{X,\bar{X}}}{s_{\rm SM}},
\end{align}
where $s_{\rm SM}$ is the entropy density of the SM (see appendix \ref{app:key_definition} for definition). We define the asymmetry $\eta$ as
\begin{align}
  \eta \equiv Y_{X} - Y_{\bar{X}},
\end{align}
and rewrite the Boltzmann equation in terms of the ratio
\begin{align}
  r
  &\equiv \frac{Y_{\bar{X}}}{Y_X},
  \, 
\end{align}
and using $x = m_X/T$ as the time variable
\begin{align}\label{eq:boltz}
  \frac{dr}{dx}
  &=
  -\frac{\lambda(x) \eta}{x^2}
  \left[ r - r_{\rm eq}(x)
  \left(\frac{1-r}
  {1-r_{\rm eq}(x)}
  \right)^2
\right].
\end{align}
Note that we write all expressions in terms of the SM temperature,
$x$. 
This formulation of the Boltzmann equation is useful when the
asymmetry $\eta\neq 0$, and becomes trivial in the $\eta=0$ limit, when 
we revert back to the equation in terms of $Y_X$.
\begin{aside}
The various quantities appearing in eq.~\eqref{eq:boltz} are defined as
follows:
\begin{align}
  \lambda(x)
  &=
  \sqrt{\frac{\pi}{45 G_{\rm N}}}
  g_*^{1/2}
  m_X
 \langle \sigma v \rangle ,
 \\
 g_*^{1/2}
 &=
  \frac{h_{\rm eff}(x)}{g_{\rm eff}^\frac12(x)}
  \left(
  1-\frac{x}{4 g_{\rm eff}(x)} \frac{dg_{\rm eff}}{dx}
  \right),
\end{align}
where $G_{\rm N}$ is Newton's constant, and where $h_{\rm eff}$ and
$g_{\rm eff}$ are the effective number of degrees of freedom for the
entropy and energy density, respectively. The values for $g_*^{1/2}$
are tabulated in ref.~\cite{Gondolo:2004sc}. The equilibrium value
$r_{\rm eq}$ appearing in eq.~\eqref{eq:boltz} is
\begin{align}
  \label{eq:req}
  r_{\rm eq}
  &=  
 \exp\left(-2\sinh^{-1}\frac{\eta}{2Y_{\rm eq}(x)}\right)
 \\
  Y_{\rm eq}(x)
  &\simeq
  \frac{45 g_X}{4\sqrt{2} \pi^{7/2} h_{\rm eff}(x)}
  x^{3/2}\xi^{3/2} e^{-x/\xi} \, .
\end{align}
where we have included the explicit factors of $\xi$.
The thermal averaged cross section is given by
\begin{align}
 \langle \sigma v \rangle (x) 
 &=
 \frac{x}{8 m_X^5 K_2^2 (x/\xi)}
 \int_{4 m_X^2}^{\infty}
 ds\, \sigma(s) \sqrt{s}(s-4m_X^2) 
 K_1\left(\frac{x \sqrt{s}}{\xi m_X}\right),
\end{align}
where $K_n(x)$ are modified Bessel functions of the second kind. 
The cross section $\sigma(s)$ includes Sommerfeld enhancement \cite{vonHarling:2014kha},
\begin{align}
  \sigma(s)
  &=
  \frac{32\pi \alpha_D^2 m_X^2}{s^2} 
  \frac{1-\frac{2m_X^2}{s}}{\sqrt{1-\frac{4m_X^2}{s}}}
  S(s,m_X)
  \, .
  \label{eq:sigmas}
\end{align}
The
thermal averaged cross section is,
\begin{align}
  \langle \sigma v \rangle (x)
  \simeq
  \frac{2 \pi \alpha_D^2}{m_X^2}
  \bar{S}_{\rm ann}.
\end{align}
where $\bar{S}_{\rm ann}$ is the thermally averaged Sommerfeld enhancement
factor. We note that the effect of $X$--$\bar{X}$ bound state formation was studied in detail in refs.~\cite{Feng:2009mn,vonHarling:2014kha} and it was found to have a subdominant effect on the relic abundance compared to the Sommerfeld enhancement. Due to the significant uncertainties associated with computing bound-state formation in detail, we neglect their contributions here.  

\end{aside}

We can solve the Boltzmann equation approximately at late times when
$r_{\rm eq} \ll r$ and match the solution at the freeze-out
temperature $x_f$. The freeze-out temperature 
is defined as the temperature where
$r$ starts deviating from $r_{\rm eq}$, such that $dr/dx, r, r_{\rm eq}$ are
all of the same order at $x_f$. Below this temperature, we can
integrate the Boltzmann equation neglecting $r_{\rm eq}$,
with the boundary condition $r(x_f) =
r_{\rm eq}(x_f)$.
Then, the final $r(x_0) \simeq r_\infty$ is
\begin{align}
  r_\infty
  &\simeq
  r_{\rm eq}(x_f)
  \exp
  \left(
  -\frac{\lambda \eta}{x_f}
  \right).
  \label{eq:rinfty11}
\end{align}
The value $x_f$ can be found by solving the following
implicit equation
\begin{align}
  \frac{dr_{\rm eq}(x_f)}{dx}
  &= 
  -\frac{\lambda(x_f) \eta}{x_f^2} r_{\rm eq}(x_f).
\end{align}
This equation can be rewritten as 
\begin{align}
  x_f
  &\simeq
  \xi_f
  \left[\log\left(\xi_f^{3/2} \lambda 
  \frac{45 g_X}{4\sqrt{2} \pi^{7/2} h_{\rm eff}(x_f)}
  \right)
  -\frac12 \log (x_f)
  +\log
  \left(
  1 + \frac16\left(\frac{\eta \lambda}{2x_f^2} \right)^2
  \right)
\right]
  \, ,
\end{align}
when 
$\eta \lambda \ll 2 x_f^2$, which is true in our parameter space of
interest.
We see then that
$\eta$ does not affect $x_f$ significantly and $x_f \sim 25 \xi_f$,
where $\xi_f \equiv \xi(T_f)$. 

Note that $\lambda$ is proportional to the annihilation cross
section, and hence is a quantity which would be 
inversely proportional to the relic abundance in the limit $\eta\to0$.
In fact, the relic comoving density in the absence of asymmetry is,
\begin{align}
  Y_{\infty}^{\eta=0} 
  \simeq
  \frac{x_f}{\lambda(x_f)}
\end{align}
so that,
\begin{align}
  r_\infty
  &\simeq
  r_{\rm eq}(x_f)
  \exp
  \left(
  -{\frac{\eta}{Y_{\infty}^{\eta=0}}}
  \right).
\end{align}
For
our parameter space, $\eta\ll Y_{eq}(x_f)$, and hence from
eq.~\eqref{eq:req} we see that  $r_{\rm eq}(x_f) \simeq 1$. 
The dark matter density today (relative to the critical density)
is
\begin{align}
  \Omega_X h^2
  &=
  \frac{m_X}{m_p}
  \frac{\Omega_B}{\eta_B}
  (Y_{X}(x_0) + Y_{\bar{X}}(x_0))
  =
  \frac{m_X}{m_p}
  \frac{\Omega_B}{\eta_B}
  \frac{1+r_\infty}{1-r_\infty} \eta
  \\&\simeq
  2.73\times10^8 \ \eta \ \left(\frac{m_X}{\mathrm{GeV}}\right)
  \frac{1+r_\infty}{1-r_\infty},\label{eq:relic_final}
\end{align}
where $\Omega_B, \eta_B$ are the baryonic energy density and
asymmetry.

\begin{figure}[tp]
  \centering
  \includegraphics[width=0.75\textwidth]{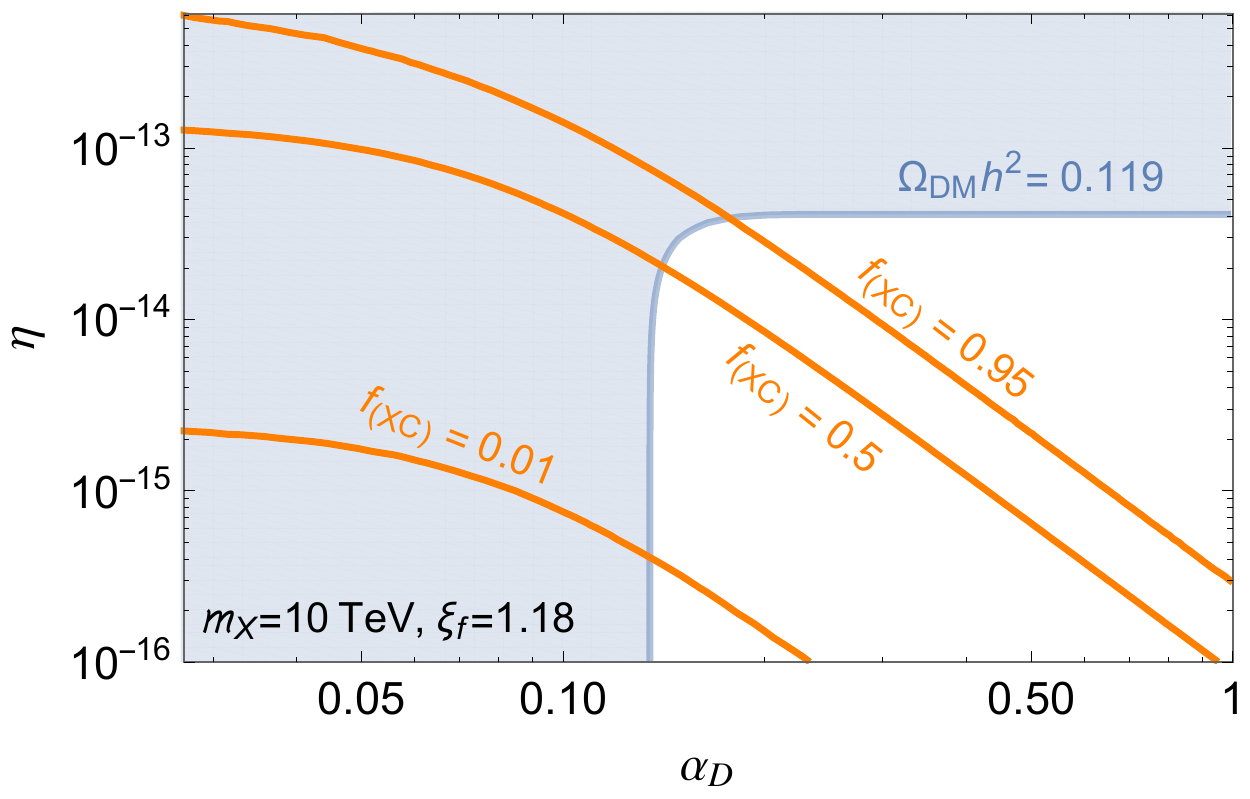}
  \caption{Relic abundance in the $XC$ dark matter scenario with both a
    symmetric as well as an asymmetric component.  The blue curve
    shows where the correct relic abundance, $\Omega_{X}h^2=0.119$ \cite{Ade:2015xua} is
    obtained, and the orange lines are contours of constant $f_{(XC)}$ fraction of mass in the asymmetric component. The shaded blue region is excluded if there are
    no other interactions that can deplete $\Omega_X$.
}
  \label{fig:asymm-symm}
\end{figure}

In figure~\ref{fig:asymm-symm} we show the region of
$\alpha_D$--$\eta$ parameter space where a significant
symmetric dark
matter relic abundance survives in the presence of an asymmetry. 
We see that there are two limiting behaviors of the solutions. When
$\eta\ll Y_{\infty}^{\eta=0}$, 
\begin{align}
  r_\infty\simeq 1-\frac{\eta}{Y_\infty^{\eta=0}}
  \Rightarrow
  \Omega_X \simeq 
  Y_\infty^{\eta=0} 
  \frac{2\Omega_B}{\eta_B}
  \frac{m_X}{m_p}
  \,.
\end{align}
In this case, the dark matter density is essentially given by the
symmetric freezeout value, and the dark matter is dominantly
symmetric. In
figure~\ref{fig:asymm-symm}, this appears as the
vertical part of the constant-$\Omega_X$ contours (i.e.~independent of
$\eta$). 
At $\eta = Y_\infty^{\eta=0}$ we see a turnover, and for
$\eta>Y_\infty^{\eta=0}$, 
\begin{align}
  r_\infty\simeq 0
  \Rightarrow
  \Omega_X \simeq \eta
  \frac{\Omega_B}{\eta_B}
  \frac{m_X}{m_p}
\end{align}
The relic abundance in this case is set by the asymmetry, and the
symmetric component is exponentially small. In this case, the
constant-$\Omega_X$ contours in 
figure~\ref{fig:asymm-symm} are horizontal (independent of $\alpha_D$).
Generally, we are interested in the parameter space where there is both a
symmetric and asymmetric population, $r_\infty\lesssim 1$,
so that we are somewhat close to the turnaround region,
\begin{align}
  \eta \lesssim Y_\infty^{\eta=0} \ .
\end{align}

At late times, the key quantity to consider is the fraction $f_{(XC)}$ of the total dark
matter mass that can end up in neutral $(XC)$ bound state. In the limit
$m_X \gg m_C$, this mass fraction is given by $f_{(XC)} \simeq (n_X -
n_{\bar{X}})/(n_X + n_{\bar{X}})$ (see eq.~\eqref{eq:f_XC_def} below), from which we obtain the relation
\begin{align}\label{eq:final_bound_state_frac}
  f_{(XC)} & \simeq \frac{1-r_\infty}{1+r_\infty} \simeq \frac{\eta}{2 Y_\infty^{(\eta = 0)} -\eta},
\end{align}
where this last relation is valid for $\eta \leq Y_\infty^{(\eta =
0)}$. Lines of constant $f_{(XC)}$ values are illustrated in figure~\ref{fig:asymm-symm}. We see that the dependence of $f_{(XC)}$ on $\eta$ is
approximately linear, so that while we do require a coincidence that
$\eta \lesssim Y_\infty^{\eta=0}$, $\mathcal{O}(1)$ variations in our
parameters do not drastically change the phenomenology. 
In fact, one
can compare this to the coincidence $\eta_B \sim Y_\infty^{\eta=0}$
present in all WIMP models.

\section{Asymmetric Recombination}\label{sec:recombination}
In this section, we compute the evolution of the number density of free $C$ particles within the dark plasma. We note that the presence of the asymmetry results in some important changes to the calculation performed in ref.~\cite{Cyr-Racine:2013ab}. 
\subsection{Notational aside}
Before diving into the details of the $(XC)$ bound-state formation, we introduce some useful notation. In the following, we denote the fraction of the total dark matter energy density that could eventually be bound into $XC$ dark atoms by $f_{(XC)}$. It is given by
\be\label{eq:f_XC_def}
f_{(XC)} \equiv \frac{\rho_C +\rho_X-\rho_{\bar{X}}}{\rho_{\rm DM}} = \frac{(m_C + m_X)(n_X - n_{\bar{X}})}{m_C(n_X - n_{\bar{X}}) + m_X (n_X + n_{\bar{X}})}\approx \frac{n_X - n_{\bar{X}}}{n_X + n_{\bar{X}}}\quad\text{for}\quad m_X\gg m_C,
\ee
where $n_i$ represents the number density of the $i^{\rm th }$ species, and where we assumed non-relativistic dark matter such that $\rho_i = n_i m_i $. In the above, $\rho_{\rm DM} = n_C m_C + m_X(n_X + n_{\bar{X}})$ is the total dark matter energy density. We note that $f_{(XC)}$ is related to the asymmetry parameter $\eta$ as given in eq.~\eqref{eq:final_bound_state_frac}. It is possible to express the number density of the different species in terms of $f_{(XC)}$ and $\rho_{\rm DM}$ 
\ba
n_X &=& \frac{\left(m_X (1+f_{(XC)}) + m_C (1-f_{(XC)})\right) \rho_{\rm DM}}{2 m_X (m_X+m_C)}\approx \frac{(1+f_{(XC)})\rho_{\rm DM}}{2 m_X},\\
n_{\bar{X}} &=& \frac{(1-f_{(XC)})\rho_{\rm DM}}{2 m_X},\\
n_C &=& n_X - n_{\bar{X}} = \frac{f_{(XC)}\rho_{\rm DM}}{m_C + m_X}\approx \frac{f_{(XC)}\rho_{\rm DM}} {m_X},
\ea
where the approximation are taken in the limit of $m_X\gg m_C$. Since we assume the $X$ and $C$ particles to be non-relativistic, we have $\rho_{\rm DM} = \Omega_{\rm DM} h^2 \tilde{\rho}_{\rm crit} (1+z)^3$, where $\tilde{\rho}_{\rm crit} = 8.098 \times 10^{-11}$ eV$^{4}$ and $z$ is redshift. To track the formation of the neutral $XC$ bound state, it is useful to introduce the fraction $R_{C}$ of  ``free'' (that is, not bound) $C$ particles in the dark sector 
\be
R_{C} \equiv \frac{n_C^{\rm free}}{n_X-n_{\bar{X}}} = \frac{n_C^{\rm free}(m_X+m_C)}{f_{(XC)} \rho_{\rm DM}}.
\ee
where $n_C^{\rm free}$ is the number density of unbound $C$ particles. Before $(XC)$ recombination, we have $R_{C}=1$. In a similar fashion, we can define the number density of ``free'' $X$ particles
\ba
n_X^{\rm free} = n_{\bar{X}} + n_C^{\rm free} = n_{\bar{X}} + (n_X-n_{\bar{X}})R_{C} &=&  \frac{\left((m_X+m_C) (1-f_{(XC)}) + 2m_X f_{(XC)} R_{C}\right) \rho_{\rm DM}}{2 m_X (m_X+m_C)}\en
&\approx&\frac{1-f_{(XC)}+2f_{(XC)}R_{C}}{2 m_X}\rho_{\rm DM},
\ea
which has the right limit as $n_X^{\rm free} \rightarrow n_X$ before dark recombination while $n_X^{\rm free} \rightarrow n_{\bar{X}}$ after bound state formation.
\subsection{Assumptions and Simplifications}
As in the case of standard hydrogen recombination \cite{Peebles:1968ja,1968ZhETF..55..278Z},  several factors  dramatically simplify the computation of the free $C$ fraction across cosmic times. We list them here:
\begin{enumerate}
\item {\bf Thermality of the dark photon bath:} Long after the dark matter freeze-out, the entropy ratio of the dark photon bath ($s_{\gamma_D}$) to that of the dark matter ($s_{\rm DM}$) is given by
\be
\frac{s_{\gamma_D}}{s_{\rm DM}} \simeq 5\times10^{12} \left(\frac{m_X}{10\,\rm TeV}\right)\frac{\xi^3}{1+f_{(XC)}}.
\ee
Unless the dark photons are unnaturally cold ($\xi\ll1$), we generally have $s_{\gamma_D}/s_{\rm DM}\gg1$ and we can thus neglect the impact of the heat transfer between the dark matter and the dark photon bath on the energy spectrum of the latter. We therefore take the dark photon bath to be exactly thermal with $f_{\gamma_D}(p) = (e^{p/T_D}-1)^{-1}$.

\item {\bf Thermal equilibrium in the dark sector:} The $X$, $\bar{X}$, and $C$ particles are in kinetic thermal equilibrium with the dark photon bath at a single temperature $T_D$ until the fraction of free $C$ particle in the plasma becomes negligible.  
 
\item {\bf No direct recombination to the ground state:} The ionization rate of $(XC)$ bound states by a dark photon at the ionization threshold in units of the Hubble rate is
\be
\frac{\Gamma_{\rm I}}{H} = \frac{ \sigma_{\rm I}  n_{(XC)} }{H} \simeq 30\left(\frac{0.1}{\alpha_D}\right)\left(\frac{m_C}{\rm MeV}\right)^{-2} \left(\frac{m_X}{10\,{\rm TeV}}\right)^{-1} f_{(XC)}(1-R_{C})(1+z),
\ee
where $\sigma_{\rm I}$ is  the cross section to ionize the ground state of the $XC$ bound state (see e.~g.~ref.~\cite{1979rpa..book.....R}), and $n_{(XC)}$ is the $(XC)$ number density. By the time the fraction of $C$ particles in bound state becomes nonnegligible (around $T_D\sim B_{(XC)}/20$), the ionization rate is always much larger than the Hubble and one can thus neglect direct recombination to the ground state. Direct recombination to the ground is inefficient since it results in the emission of a dark photon that immediately ionizes another $(XC)$ bound state, hence yielding no net recombination. 
\end{enumerate}
With the above conditions satisfied, $(XC)$ bound state formation proceeds similarly to the standard hydrogen recombination \cite{Peebles:1968ja,1968ZhETF..55..278Z,Seager:1999km}, with the main (and very important) difference being the presence of a symmetric population of $X$ and $\bar{X}$.

\subsection{Evolution of free $C$ fraction}
The evolution equation for $ R_C$ is
\be\label{eq:ionization_evol}
\frac{d R_{C}}{dz} = -\frac{\langle \sigma_{\rm rec}^{(2)} v\rangle C_{\rm Peebles}}{H(z) (1+z)}\left[\left(\frac{m_C T_D}{2\pi}\right)^{3/2}e^{-B_{(XC)}/T_D}(1- R_{C}) -  R_{C}\left(n_{\bar{X}}+(n_X - n_{\bar{X}}) R_{C}\right)\right],
\ee
where $\langle \sigma_{\rm rec}^{(2)} v\rangle$ is the net volumetric recombination rate to the $n=2$ atomic state, $B_{(XC)} = \alpha_D^2 m_C/2$ is the binding energy, $T_D$ is the temperature of the dark photons, and $C_{\rm Peebles}$ is the Peebles factor \cite{Peebles:1968ja, Seager:1999km} described below. The principal difference between the standard symmetric recombination equation \cite{Seager:1999km,Dodelson-Cosmology-2003} and eq.~\eqref{eq:ionization_evol} is the presence of the term proportional to $R_{C} n_{\bar{X}}$ on the right-hand side. This linear term in $R_{C}$ arises because there are many available $X$ particles for each $C$ looking to recombine, even when a sizable population of $(XC)$ bound states has already formed. This is in stark contrast with standard symmetric recombination where the number of available free protons is rapidly depleted as hydrogen atom formation occurs.  As long as the recombination rate $n_X\langle \sigma_{\rm rec}^{(2)} v\rangle$ is larger than the Hubble rate, the large population of free $X$ particles in our model tend to make $(XC)$ bound state formation extremely efficient, leading to an exponentially suppressed abundance of free $C$ particles. On the other hand, models for which the recombination rate is smaller than the Hubble rate for $T_D < B_{(XC)}$ experience no bound state formation at all, and have $R_{C}\sim1$ at late times.  An interesting consequence of this analysis is that the atomic physics details encoded in the Pebbles factor matter very little for a large swath of the parameter space. Only models for which $(\langle \sigma_{\rm rec}^{(2)} v\rangle n_X/H)|_{z=z_{\rm rec}}\sim1$ are sensitive to the $(XC)$ atomic physics. This makes our prediction for the free $C$ fraction particularly robust, and obviates the need for a detailed treatment of dark recombination such as that presented in ref.~\cite{Cyr-Racine:2013ab}. It is thus sufficient to follow the simple approach presented in refs.~\cite{Peebles:1968ja,Dodelson-Cosmology-2003}, which is based on eq.~\eqref{eq:ionization_evol} together with the rates given below.

\begin{figure}[t]
\begin{center}
\includegraphics[width=0.65\textwidth]{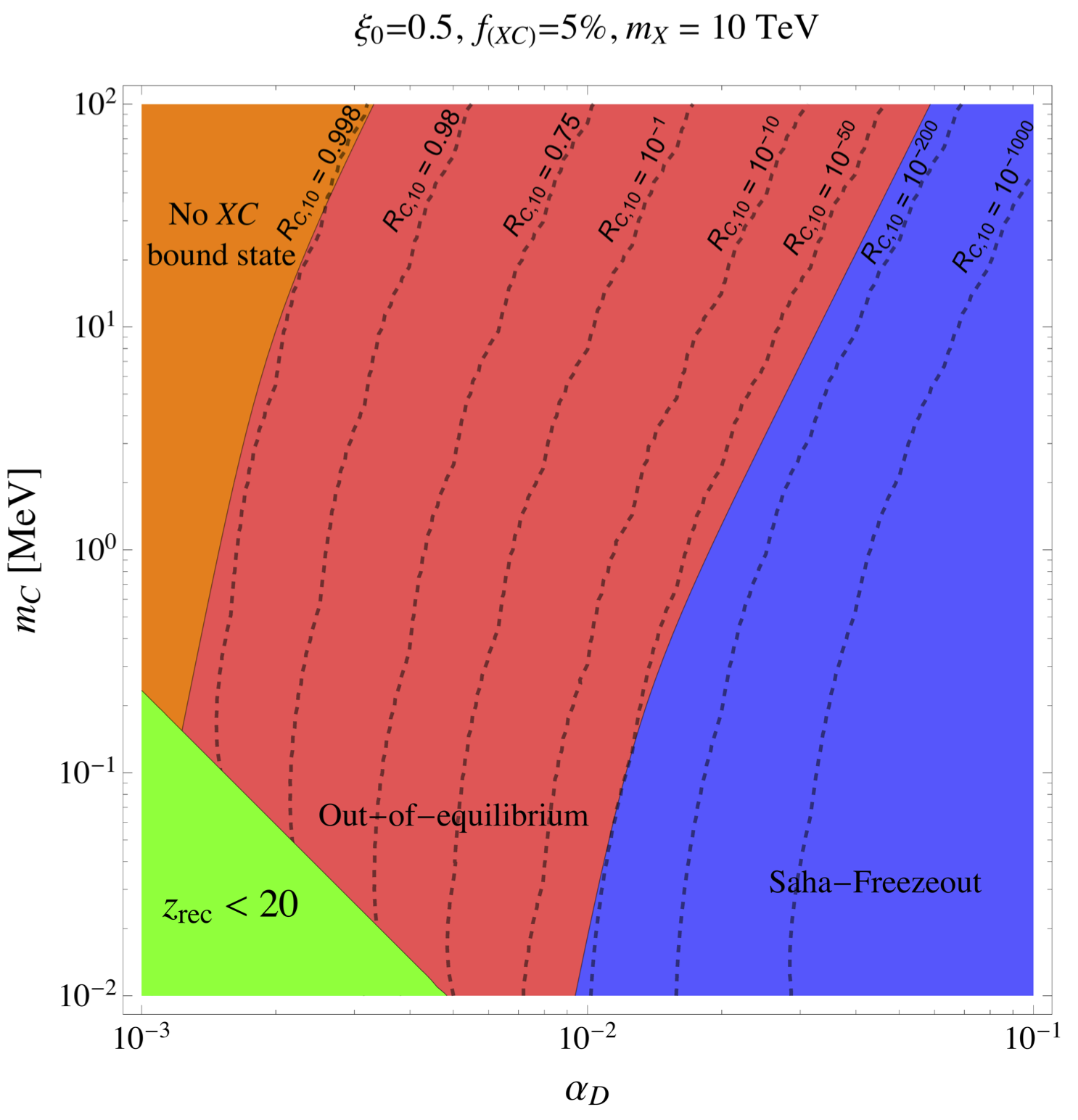}
\caption{Fraction of free $C$ particles at late times as a function of $\alpha_D$ and $m_C$. The dashed lines indicate contours of constant free $C$ fraction at $z=10$, $R_{C,10}\equiv R_C(10)$. The slight wobble in the contours is caused by the numerical interpolation scheme used. The blue region labelled ``Saha-Freezeout'' indicate where the solution given in eq.~\eqref{eq:freezeout_soln_XC} applies. The line delimitating this region is given by eq.~\eqref{eq:saha_eq_region}. The red region labelled ``Out-of-equilibrium'' indicates where a full numerical solution is necessary, and the orange region labelled ``No $XC$ bound state'' denotes where dark recombination does not take place. The line delimitating this region from the red region is given by eq.~\eqref{eq:no_bound_state}. In the green region, recombination occurs too late ($z_{\rm rec} < 20$) to be accurately described by eq.~\eqref{eq:ionization_evol}. Here, we fix $\xi_0 = 0.5$, $m_X = 10$ TeV, and $f_{(XC)} = 0.05$. For these parameters, we need $\alpha_D\sim 0.15 $ in order to obtain the right dark matter relic abundance implying that recombination is well described by the Saha-Freezeout regime for $m_C$ of order MeV. }
\label{Fig:ionization}
\end{center}
\end{figure}

The recombination rate appearing in eq.~\eqref{eq:ionization_evol} is approximately given by \cite{1951ApJ...114..407S,Dodelson-Cosmology-2003}
\be
\langle \sigma_{\rm rec}^{(2)} v\rangle \approx 0.448\frac{64 \pi}{\sqrt{27\pi}} \frac{\alpha_D^2}{m_C^2}\left(\frac{B_{(XC)}}{T_D}\right)^{1/2} \ln{\left(\frac{B_{(XC)}}{T_D}\right)},
\ee
which is valid for $T_D < B_{(XC)}$, that is, the relevant range for bound state formation. Since direct recombination to the ground state is inefficient, bound state formation has to proceed through either the $2s$ or $2p$ states. The singlet state can decay to the ground state only via the 2-dark-photon transition. The triplet state can directly decay to the ground state via spontaneous emission of a dark Lyman-$\alpha$ photon. As in the case of standard hydrogen recombination, the $2p$ to $1s$ transition results in a net bound state formation only if the dark Lyman-$\alpha$ photon can redshift out of the Lyman-$\alpha$ line before it is reabsorbed by another $(XC)$ bound state. This is encoded in the so-called ``Peebles'' factor $C_{\rm Peebles}$ which is given by
\be 
C_{\rm Peebles} = \frac{R_{{\rm Ly}\alpha} + \Lambda_{2\tilde{\gamma}}}{R_{{\rm Ly}\alpha} + \Lambda_{2\tilde{\gamma}}+\beta^{(2)}},
\ee
where $R_{{\rm Ly}\alpha}$ is the rate for dark Lyman-$\alpha$ photons to redshift out of the line, $\Lambda_{2\tilde{\gamma}}$ is the $2s$ to $1s$ dark two-photon decay rate, and $\beta^{(2)}$ is the photoionization rate for the $n=2$ state. These rates are given in ref.~\cite{Cyr-Racine:2013ab}, but we use here the simpler results from ref.~\cite{Dodelson-Cosmology-2003},
\be
R_{{\rm Ly}\alpha} \approx \frac{3^3 \alpha_D^6 m_C^3 H(z)(1+z)}{8(8\pi)^2n_C(1- R_{C})},\qquad \Lambda_{2\tilde{\gamma}} \approx 2.0\times 10^{10} \left(\frac{\alpha_D}{0.1}\right)^8\left( \frac{m_C}{\rm MeV}\right)\, {\rm s}^{-1},
\ee
\be
 \beta^{(2)} = \langle \sigma_{\rm rec}^{(2)} v\rangle \left(\frac{m_C T_D}{2\pi}\right)^{3/2}e^{-B_{(XC)}/(4T_D)}.
\ee
We now have all the ingredients to solve eq.~\eqref{eq:ionization_evol}. To a good approximation, dark recombination occurs once the temperature of the dark sector has fallen to $T_D\sim 0.02 B_{(XC)}$ \cite{Cyr-Racine:2013ab}, which corresponds to a recombination redshift 
\be
z_{\rm rec} \simeq 8.9\times 10^5 \left(\frac{\alpha_D}{0.1}\right)^2\left(\frac{m_C}{{\rm MeV}}\right)\left(\frac{0.5}{\xi}\right).
\ee
In figure~\ref{Fig:ionization} we illustrate the free $C$ fraction at late times as a function of $\alpha_D$ and $m_C$ for a fiducial model with $m_X = 10$ TeV, $f_{(XC)} = 5\%$, and $\xi_0 = 0.5$. The dashed lines indicate contours of constant free $C$ fraction at redshift $z=10$. The different colored regions indicate the different types of solution to eq.~\eqref{eq:ionization_evol}, which we now consider in more detail.
%

\subsubsection{Saha-Freezeout regime}
Whenever the factor multiplying the square bracket in eq.~\eqref{eq:ionization_evol} is large, dark recombination will proceed in equilibrium and will be described by the Saha equation, which in this case reads
\be\label{eq:saha}
\frac{ R^{\rm eq}_C(n_{\bar{X}} + n_C  R^{\rm eq}_C)}{1 -R^{\rm eq}_C} = \left(\frac{m_C T_D}{2\pi}\right)^{3/2} e^{-B_{(XC)}/T_D}
\ee
As explained above, the presence of the symmetric $X$ component can have a dramatic effect on the residual fraction of free $C$ particles at late times. For $f_{(XC)}\ll1$, bound state formation tends to be very efficient since there are $\sim1/(2f_{(XC)})$ available $X$ particles in the bath for every $C$ particle that is looking to recombine. As long as the recombination rate obeys $\langle \sigma_{\rm rec}^{(2)} v\rangle n_{X}/H  \gg1 $, bound state formation will thus proceed in equilibrium and the free $C$ particle fraction is well approximated by eq.~\eqref{eq:saha}. The dark atomic physics encoded in the Peebles factor essentially plays no role here. In this regime, the residual free $C$ fraction at late times is exponentially suppressed with
\be
R^{\rm eq}_C(z)\simeq \frac{1}{n_{\bar{X}}}\left(\frac{m_C T_D}{2\pi}\right)^{3/2}  e^{-B_{(XC)}/T_D}\qquad z_{f}<z\ll z_{\rm rec},
\ee
where $z_f$ is the redshift at which the recombination process goes out of equilibirum, which is roughly given by the condition $(\langle \sigma_{\rm rec}^{(2)} v\rangle n_{X}/H)|_{z_f}  \sim1 $. The actual asymptotic value of $R_{C}$ at late times can be computed using a freezeout calculation similar\footnote{We note that eq.~\eqref{eq:boltz} can be recast in the form of eq.~\eqref{eq:ionization_evol} with the transformation
\be
\eta \to \frac{1-f_{(XC)}}{2 f_{(XC)}},\qquad r\to \frac{2 f_{(XC)}R_{C}}{1+ f_{(XC)}(2R_{C}-1)} \nonumber. 
\ee
 } to that described in section \ref{sec:relic-abundance}, but taking into account the different redshift dependence of the recombination and Hubble rates. A detailed calculation similar to that leading to eq.~\eqref{eq:rinfty11} above gives
 \begin{align}\label{eq:freezeout_soln_XC}
 R_{C}(z) &\simeq R^{\rm eq}_C(z_f)\exp{\left[-\frac{\lambda_D}{\omega} (F(z) - F(z_f)) \right]}\qquad z \ll z_f,
 \end{align}
where
\be
F(z) = 4 \tanh^{-1}\left(\sqrt{ 1+\omega(1+z)}\right) - 2\sqrt{ 1+\omega(1+z)}\left( 2 + \ln{\left(\frac{A_D}{1+z}\right)}  \right),
\ee
\be
\omega = \frac{\Omega_{\rm rad}}{\Omega_{\rm m}},\qquad A_D \simeq 2.1\times 10^9 \frac{\alpha_D^2}{ \xi }\left(\frac{m_C}{\rm MeV}\right),
\ee
\be
 \lambda_D \simeq \frac{10^{2}\alpha_D^3(1-f_{(XC)})}{h\sqrt{\Omega_{\rm m}\xi}}\left(\frac{\Omega_{\rm DM}h^2}{0.12}\right)\left(\frac{m_C}{\rm MeV}\right)^{-3/2}\left(\frac{m_X}{10\,{\rm TeV}}\right)^{-1},
\ee
where $\Omega_{\rm rad}$ and $\Omega_{\rm m}$ are the total energy density in radiation (including neutrinos and photons) and matter, respectively, both expressed in units of the critical density of the Universe. The recombination freezeout redshift can be estimated by solving the implicit equation
\be
\frac{\lambda_D (1+z_f)^2 (\ln{A_D}-\ln{(1+z_f)})}{\sqrt{ 1+\omega(1+z_f)}} = A_D - \frac{3}{2}(1+z_f).
\ee
The validity of the above freezeout solution can be verified by comparing it with the exact numerical solution to eq.~\eqref{eq:ionization_evol}. We determine that it is generally valid when $(\langle \sigma_{\rm rec}^{(2)} v\rangle n_X/H)|_{z=z_{\rm rec}}\gtrsim10^2$, which translate to a parameter range given by
\be\label{eq:saha_eq_region}
\frac{\alpha_D^4 (1 + f_{(XC)})}{\xi} \left(\frac{m_C}{\rm MeV}\right)^{-1} \left(\frac{m_X}{10\,{\rm TeV}}\right)^{-1}\left(\frac{\Omega_{\rm DM}h^2}{0.12}\right)\gtrsim  10^{-7}.
\ee
The parameter range is indicated in figure~\ref{Fig:ionization} by the blue region. 

\subsubsection{Nonequilibrium regime}
The evolution of models with $ 10^{-3} \lesssim (\langle \sigma_{\rm rec}^{(2)} v\rangle n_X/H)|_{z_{\rm rec}} \lesssim 10^2 $ significantly deviate from the equilibrium Saha solution as soon as the recombination process begins. Here, the details of the dark atomic physics can make a significant difference on the asymptotic value of the free $C$ fraction. The Peebles factor can play an important in this regime, and one must numerically solve eq.~\eqref{eq:ionization_evol} to obtain an accurate value of $R_{C}$ at late times. The parameters in this regime satisfy
\be
10^{-12} \lesssim \frac{\alpha_D^4 (1 + f_{(XC)})}{\xi} \left(\frac{m_C}{\rm MeV}\right)^{-1} \left(\frac{m_X}{10\,{\rm TeV}}\right)^{-1}\left(\frac{\Omega_{\rm DM}h^2}{0.12}\right)<  10^{-7},
\ee
which is indicated in figure~\ref{Fig:ionization} by the red region.

\subsubsection{No bound state formation}
%
Finally, models for which $ (\langle \sigma_{\rm rec}^{(2)} v\rangle n_X/H)|_{z_{\rm rec}} \lesssim 10^{-3} $ form a negligible number of $XC$ bound states. This regime applies when
\be\label{eq:no_bound_state}
 \frac{\alpha_D^4 (1 + f_{(XC)})}{\xi} \left(\frac{m_C}{\rm MeV}\right)^{-1} \left(\frac{m_X}{10\,{\rm TeV}}\right)^{-1}\left(\frac{\Omega_{\rm DM}h^2}{0.12}\right)<  10^{-12},
\ee
which is indicated by the orange region in figure~\ref{Fig:ionization}. In this regime, less than $0.2\%$ of $C$ particles end up in neutral bound states.
\section{Structure Formation Constraints}\label{sec:structure_formation}
We now turn our attention to the cosmological growth of dark matter density fluctuations in our darkly charged model. Compared to a purely symmetric dark charged scenario \cite{Feng:2009mn,Ackerman:2008gi}, the main new element in our scenario is the presence of the light $C$ particle catalyzing the net momentum transfer rate between the dark radiation and the heavy $X$ particles, which can significantly delay the epoch of the dark matter kinetic decoupling and the onset of the growth of structure. This leads to a suppression of the amplitude of the matter power spectrum on scales $k > k_{\rm dec}$ that enter the causal horizon before dark matter kinetic decoupling. Observationally, a late epoch of kinetic decoupling leads to a reduced abundance of small-mass subhalos \cite{Boehm:2000gq,Boehm:2001hm,Boehm:2004th,Aarssen:2012fx,Cyr-Racine:2013ab,Schewtschenko:2014fca,Buckley:2014hja,2016MNRAS.460.1399V,Schewtschenko:2015rno} within the Local Group. In addition, the delayed formation of structure on scales $k\sim k_{\rm dec}$ leads to satellite galaxies that are less centrally concentrated, resulting in shallower rotation curves for these objects. The presence of significant dark matter self-interaction in our model could amplify this latter effect via the formation of large central cores within dwarf galaxies.

Detailed simulations \cite{2016MNRAS.460.1399V,Schewtschenko:2015rno} of Milky Way-like dark matter halos show that suppressing the linear matter power spectrum by a factor of $\sim2$ on scales $k \lesssim 10h/$Mpc very likely results in a population of dwarf satellites that is incompatible with observations, both in terms of the overall abundance of massive subhalos \cite{2011PhRvD..83d3506P} and from their internal structure \cite{2010MNRAS.406.1220W}. To constrain our model, we thus require that the criterion\footnote{In this work, we neglect the small difference between the wavenumber of kinetic decoupling $k_{\rm dec}$ and the wavenumber $k_{1/2}$ at which the matter power spectrum is suppressed by a factor of 2 compared to CDM.}  $k_{\rm dec} > 10h/$Mpc be satisfied in order for an $XC$ dark matter model to be viable. We note that since the self-interaction strengths used in the simulations of ref.~\cite{2016MNRAS.460.1399V} were significantly smaller than the typical values present in our model (due their use of a massive mediator instead of a massless dark photon), the above bound is likely a \emph{necessary} but not \emph{sufficient} condition for an $XC$ model to be allowed by observations. We caution however that detailed simulations would have to be performed to assess the viability of model with $k_{\rm dec} \sim 10h/$Mpc, especially since large dark matter self-interaction cross section can lead to an unexpected phenomenology (see discussion in ref.~\cite{Agrawal:2016quu}). For a Planck 2015 cosmology \cite{Ade:2015xua} and using the relation between conformal time and redshift given in eq.~\eqref{eq:app_tau_to_z}, the above criterion is equivalent to demanding that the redshift of dark matter kinetic decoupling obeys $z_{\rm dec} > 10^6$.

Measurement of the  Lyman-$\alpha$ absorption spectrum \cite{2006PhRvL..97s1303S,2013PhRvD..88d3502V,2015JCAP...11..011P,2016JCAP...08..012B} also provides a bound on the epoch of dark matter kinetic decoupling. This constraint is usually phrased in term of the allowed values for the mass of a warm dark matter particle. For instance, ref.~\cite{2013PhRvD..88d3502V} finds this latter constraint to be $m_{\rm WDM} > 3.3$ keV at $95\%$ confidence level, while ref.~\cite{2016JCAP...08..012B} finds $m_{\rm WDM} > 4.1$ keV at $95\%$ confidence level. These bounds approximately translate to $k_{\rm dec} \gtrsim 24h/$Mpc and $k_{\rm dec} \gtrsim 31h/$Mpc, respectively. While more stringent than those derived from dwarf galaxies in the Local Group, these constraints are subject to a number of mitigating factors related to reionization and the modeling of the interstellar gas. For definitiveness, we adopt in this work the weaker but more reliable criterion $z_{\rm dec} > 10^6$, but note that the current Lyman-$\alpha$ measurements can only strengthen this bound by a factor of at most $\sim3$.  

In this section, we first present the equations of motion for the different dark matter components and argue that they can be simplified to two ``effective'' equations describing the evolution of dark matter density perturbations and bulk velocity. Using this simplified picture, we apply the bound on the redshift of kinetic decoupling discussed above in order to determine the allowed parameter space of $XC$ dark matter. We note in passing that $X$-$\bar{X}$ bound state formation plays essentially no role in the cosmological evolution of linear dark matter density fluctuations.
\subsection{Evolution equations}
The key equations governing the growth of dark matter fluctuations in the presence of new interactions have been worked out in detail in ref.~\cite{Cyr-Racine:2015ihg}. A summary of the derivation with a focus on the model under consideration is given in appendix \ref{app:DM_flucts}. As described there, the linear equations for the dark matter density and bulk velocity fluctuations are
\begin{align}
\dot{\de}_i +\theta_i -3\dot{\phi} &= 0, \label{eq:density_fluct_i} \\
\dot{\theta}_i-c_{i}^2k^2\de_i+\mathcal{H} \theta_i- k^2\psi  &=  \dot{\kappa}_{i\gamma_D}\left(\theta_i - \theta_{\gamma_D}\right) + \sum_{j\neq i} \dot{\kappa}_{ij}(\theta_i - \theta_j),\label{eq:bulk_vel_i}
\end{align}
where $\{i,j\}\in X, \bar{X}, C$. Here, an overhead dot denotes a derivative with respect to conformal time, $\de_i \equiv \de n_i/\bar{n}_i$ is the number density contrast for species $i$, with $\bar{n}_i$ being the homogeneous and isotropic part of the number density, $\theta_i \equiv i \kk\cdot \vec{v}_i$ is the divergence of the bulk velocity (in Fourier space), $\phi$ and $\psi$ are the two gravitational potentials in Conformal Newtonian gauge \cite{Ma:1995ey}, $c_i$ is the adiabatic sound speed of species $i$, $\mathcal{H}$ is the conformal Hubble rate, $k = |\kk|$ is the magnitude of the Fourier wavenumber, and $\dot{\kappa}_{ij}$ is the opacity (momentum transfer rate) for species $i$ to scatter off species $j$. In terms of the distribution function $f_i(\pp{})$ of species $i$, the number density and bulk velocity are given by
\be
n_i = \bar{n}_i(1+\de_i) = g_i\int\frac{d^3p}{(2\pi)^3}f_i (\pp{}), \qquad  \vec{v}_i =\frac{g_i}{\bar{n}_i}\int\frac{d^3p}{(2\pi)^3}f_i(\pp{})\frac{\pp{}}{(p^2+m_i^2)^{1/2}}, 
\ee
where $g_i$ is the spin degeneracy of species $i$. There is a related set of equations describing the evolution of the dark photon fluctuations (see e.g. ref.~\cite{Cyr-Racine:2015ihg}), but since we are mainly interested in dark matter fluctuations we don't consider them here. We note that for dark radiation, the bulk velocity should be interpreted as the net heat flux from hotter to colder regions of the dark plasma.

\begin{figure}[h]
\begin{center}
\includegraphics[width=0.8\textwidth]{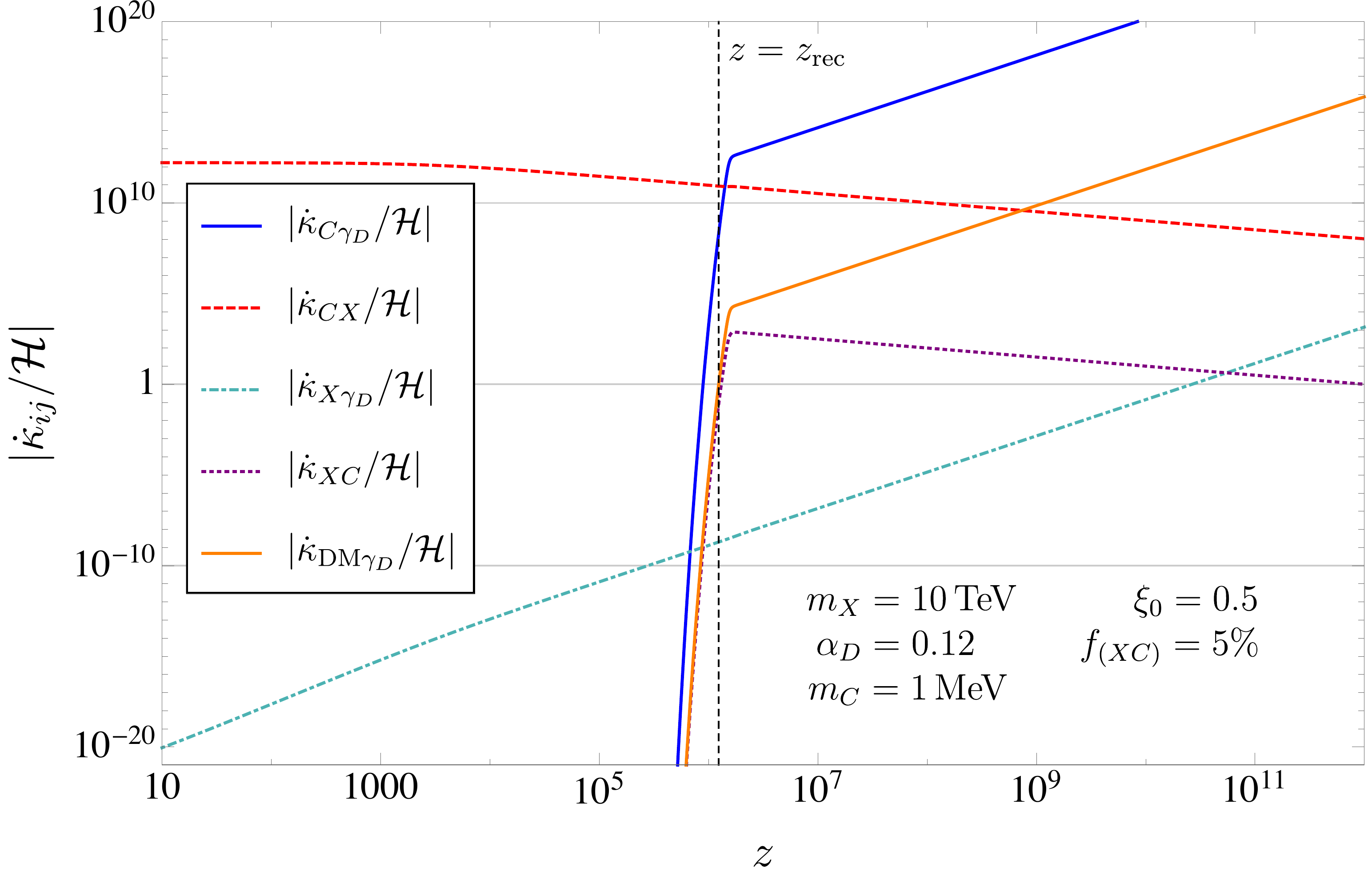}
\caption{Scattering opacities divided by the conformal Hubble expansion rate for the relevant scattering channels described in Eqs.~\eqref{eq:XC_gamma_scat}-\eqref{eq:gamma_barX_scat} and \eqref{eq:XC_scat}. We illustrate a model with $m_X = 10$ TeV, $m_C = 1$ MeV, $\alpha_D = 0.12$, $\xi_0 = 0.5$ and $f_{(XC)} = 0.05$. }
\label{Fig:opacity}
\end{center}
\end{figure}

There are two types of opacities appearing on the right-hand side of eq.~\eqref{eq:bulk_vel_i}: Thomson opacities describing scattering between dark matter particles and dark photons, and Coulomb opacities capturing the scattering between darkly-charged particles. We list in appendices \ref{app:sec_Thomson} and \ref{app:sec_Coulomb} the relevant expressions for these opacities, which represent the rate at which momentum is transferred between the different constituents. The ratios of these opacities to the conformal Hubble rate are illustrated in figure~\ref{Fig:opacity} for parameters of interest. Whenever $\dot{\kappa}_{ij}/\mathcal{H} >1$, the $i$ and $j$ species are in kinetic equilibrium with one another, implying that their bulk velocity is nearly identical with $\theta_i\simeq\theta_j$. Before $XC$ bound state formation, the situation in the dark plasma is thus as follows. The light $C$ particles are kept in kinetic equilibrium with the dark photons through frequent Thomson scattering. By contrast, the heavy $X$ and $\bar{X}$ particles cease to directly scatter off the dark photons early on since their Thomson cross section is suppressed by a factor of $(m_C/m_X)^2\ll1$ compared to that of the $C$ particles. The heavy dark matter particles are nonetheless kept in kinetic equilibrium with the rest of the dark plasma through Coulomb scattering with the light $C$ particles. This is reminiscent of the standard baryon-photon plasma in which the heavy proton and helium nuclei are kept in kinetic equilibrium with the radiation via Coulomb scattering with electrons, even though direct scattering between nuclei and photons is strongly suppressed. 

In this tightly-coupled regime where the whole dark sector behaves as a single fluid, it is possible to simplify the system of equations given in eqs.~\eqref{eq:density_fluct_i} and \eqref{eq:bulk_vel_i} to two effective equations describing the evolution of the \emph{total} dark matter density and bulk velocity perturbations, denoted by $\de_{\rm DM}$ and $\theta_{\rm DM}$, respectively. Explicitly, we have
\be
\de_{\rm DM} = \frac{1+f_{(XC)}}{2}\de_X + \frac{1-f_{(XC)}}{2}\de_{\bar{X}} + \frac{m_C f_{(XC)}}{m_X + m_C}\de_C,
\ee
and a similar expression holds for $\theta_{\rm DM}$. Since we are working in the limit $m_C\ll m_X$ and we have $\de_X\simeq \de_{\bar{X}}$ since both obey very similar equations of motion, the total dark matter density perturbation is simply $\de_{\rm DM}\approx \de_X$. The effective dark matter equations take the form
\begin{align}
\dot{\de}_{\rm DM} +\theta_{\rm DM} -3\dot{\phi} &= 0, \label{eq:density_fluct_DM} \\
\dot{\theta}_{\rm DM}-c_{\rm DM}^2k^2\de_{\rm DM}+\mathcal{H} \theta_{\rm DM}- k^2\psi  &=  \dot{\kappa}_{{\rm DM}\gamma_D}\left(\theta_{\rm DM} - \theta_{\gamma_D}\right),\label{eq:bulk_vel_DM}
\end{align}
where the dark matter opacity is given by
\be
\dot{\kappa}_{{\rm DM}\gamma_D} = -\frac{4}{3}(\Omega_{\gamma}h^2) \tilde{\rho}_{\rm crit}  \frac{8\pi\alpha_D^2}{3 m_C^2} \frac{\xi_0^4 f_{(XC)}R_{C}(z)}{m_C+m_X}(1+z)^3,
\ee
where $\Omega_\gamma h^2 = 2.47 \times 10^{-5}$ is the physical density of the CMB photons today, in units of the critical density of the Universe. The parametric dependence of $\dot{\kappa}_{{\rm DM}\gamma_D}$ is rather intuitive with $8\pi\alpha_D^2/(3m_C^2)$ being the Thomson cross section the $C-\gamma_D$ scattering, $\xi_0^4$ providing the scaling with the dark photon energy density, $f_{(XC)}R_C$ controls the fraction of scatterers that can efficiently exchange momentum with the dark radiation, and the factor $1/m_X$ arises because the dark radiation must push around the heavy $X$ and $\bar{X}$ in order to maintain kinetic equilibrium within the dark plasma. 
%

\subsection{Dark matter kinetic decoupling}
\begin{figure}[t!]
\begin{center}
\includegraphics[width=0.75\textwidth]{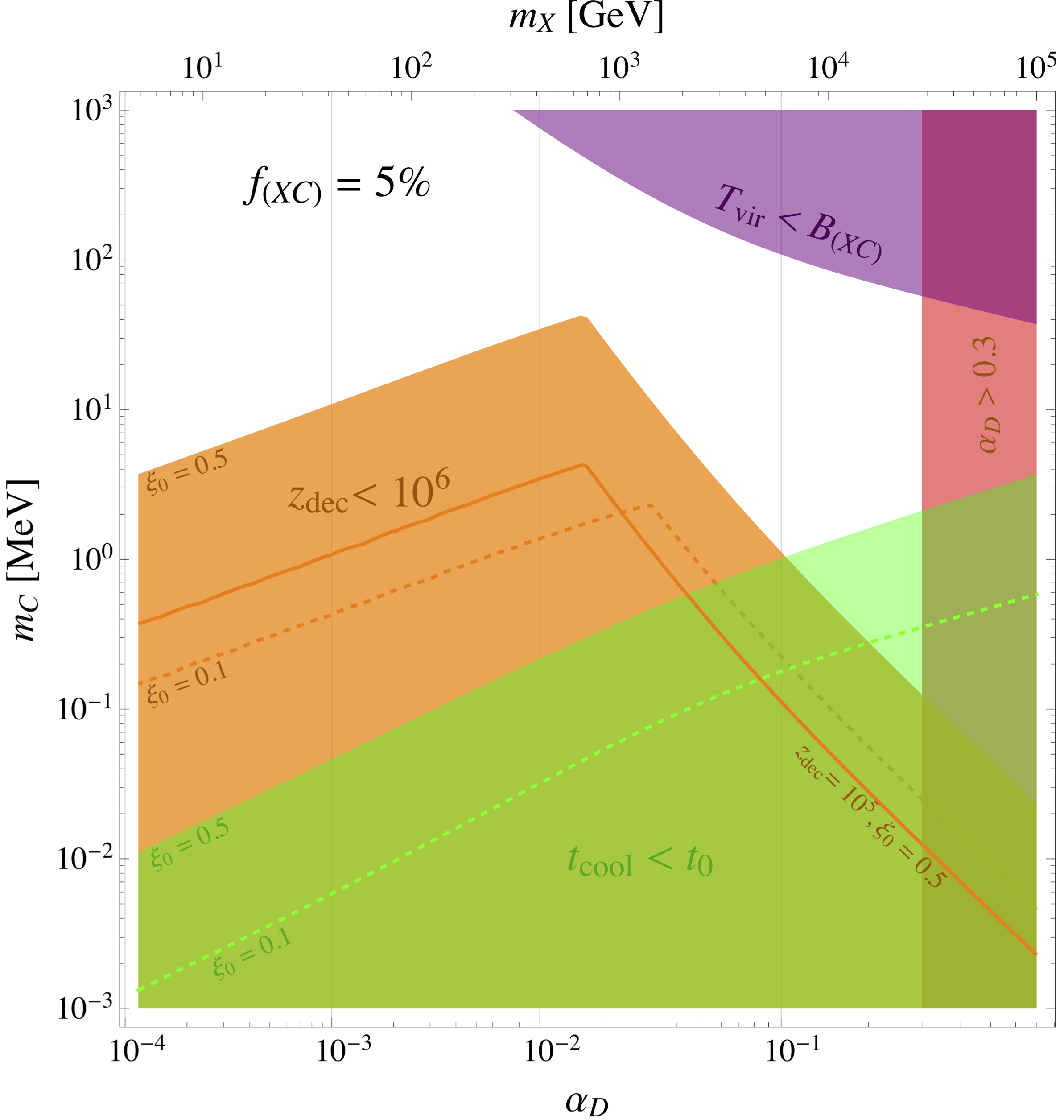}
\caption{Parameter space for $XC$ dark matter. In this plot, we assume that the relation between $\alpha_D$ and $m_X$ is fixed by demanding that $\Omega_Xh^2 = 0.119$ (see eq.~\eqref{eq:relic_final}). The top axis shows the value of the $X$ particle mass corresponding to each value of $\alpha_D$. In the orange regions, dark matter kinematically decouples from the dark photons too late, leading to a suppression of small-scale structures that is in tension with observations of Local Group dwarf galaxies. The larger orange region denotes this latter constraint when $\xi_0 = 0.5$, while the orange dashed line shows what the kinetic decoupling constraint looks like when $\xi_0 = 0.1$. The solid orange line indicates the constrained region if the kinetic decoupling bound is relaxed to $z_{\rm dec} > 10^5$ when $\xi_0 = 0.5$. The red region denoted $\alpha_D > 0.3$ illustrates the parameter space where non-perturbative effects are most likely important. The green regions denote the parameter combinations where the $C$ particles can shed an $\mathcal{O}(1)$ fraction of their kinetic energy through either Compton or Bremsstrahlung cooling within the age of the Universe. We emphasize that this region is not necessarily ruled out, but rather indicates where dissipative dynamics can play an important role shaping the internal structure of galaxy-scale dark matter halos. Again, the large green-shaded region is valid for $\xi_0=0.5$ while the smaller region bounded by the dashed green line is for $\xi_0 = 0.1$. The purple region shows the parameters for which the typical virial temperature of a Milky Way-size halo is too low to ionize the $(XC)$ bound states.}
\label{fig:constraints}
\end{center}
\end{figure}

We now turn our attention to the epoch at which dark matter kinematically decouples from the dark photon bath and starts forming structures. Using the general criterion $\dot{\kappa}_{{\rm DM}\gamma_D}(z_{\rm dec})=\mathcal{H}(z_{\rm dec})$ to solve for the redshift $z_{\rm dec}$ at which dark matter kinetic decoupling occurs, we identify two regimes of interest:
\begin{enumerate}
\item $z_{\rm dec} > z_{\rm rec}$: In this regime, kinetic decoupling occurs before $(XC)$ bound state formation, and we can thus solve analytically the condition $\dot{\kappa}_{{\rm DM}\gamma_D}(z_{\rm dec})=\mathcal{H}(z_{\rm dec})$ since $R_{C}(z_{\rm dec})\simeq1$. We obtain
\be
z_{\rm dec} \simeq 1.4\times 10^4 \left(\frac{m_C}{{\rm MeV}}\right)\left(\frac{0.1}{\alpha_D}\right)\sqrt{\frac{m_X+m_C}{10\,\rm TeV}}\left(\frac{0.5}{\xi_{\rm dec}}\right)^2\sqrt{\frac{0.05}{f_{(XC)}}},
\ee
provided that the condition 
\be
\alpha_D^6 \xi_{\rm dec}^2 f_{(XC)} \left(\frac{m_X}{10\,\rm TeV}\right)^{-1} < 3\times10^{-12}
\ee
is satisfied. This last condition comes from demanding that $z_{\rm dec} > z_{\rm rec}$. Here, $\xi_{\rm dec} \equiv \xi(z_{\rm dec})$. We note that this  generally occurs in the weakly-coupled regime ($\alpha_D\ll1$), or for a cold dark sector ($\xi_{\rm dec}\ll1$).

\item $z_{\rm dec} \simeq z_{\rm rec}$: In this case, the precipitous drop in the free $C$ fraction at $z\simeq z_{\rm rec}$ entirely controls kinetic decoupling. In this case, we simply have
\be
z_{\rm dec} \simeq 8.9\times 10^5 \left(\frac{\alpha_D}{0.1}\right)^2\left(\frac{m_C}{{\rm MeV}}\right)\left(\frac{0.5}{\xi_{\rm dec}}\right),
\ee
provided that we have
\be
\alpha_D^6 \xi_{\rm dec}^2 f_{(XC)} \left(\frac{m_X}{10\,\rm TeV}\right)^{-1} > 3\times10^{-12}.
\ee
\end{enumerate}
We illustrate in figs. \ref{fig:constraints} and \ref{fig:constraints_fixed_mX} the parameter space excluded by demanding that $z_{\rm dec} > 10^6$. Specifically, the orange-shaded regions denote the parameter combinations that are ruled out by structure formation constraints. 

It is important to emphasize that the $XC$ model has other structure formation-related features that can help distinguishing it from, say, warm dark matter. On scales entering the causal horizon before the epoch of kinetic decoupling, dark acoustic oscillations generally lead to important changes to the subhalo mass function on small scales \cite{Buckley:2014hja,2016MNRAS.460.1399V}. For instance, $XC$ dark matter generally predicts a shallower suppression of the mass function at small masses compared to an equivalent warm dark matter model (see Fig.~5 of ref.~\cite{Buckley:2014hja}). Another important distinguishing feature is the redshift evolution of the matter power spectrum \cite{Boehm:2003xr}. Indeed, the impact of dark acoustic oscillations on matter clustering becomes more important at higher redshift since nonlinearities in the density field had less time to erase them. As galaxy and weak lensing surveys begin measuring matter clustering at increasing redshifts, one could search for oscillations in the matter power spectrum that could indicate that dark matter was coupled to a relativistic species at early times.  

%

\section{Dark photons and Hubble rate}\label{sec:Hubble}
\begin{figure}[h]
\begin{center}
\includegraphics[width=0.65\textwidth]{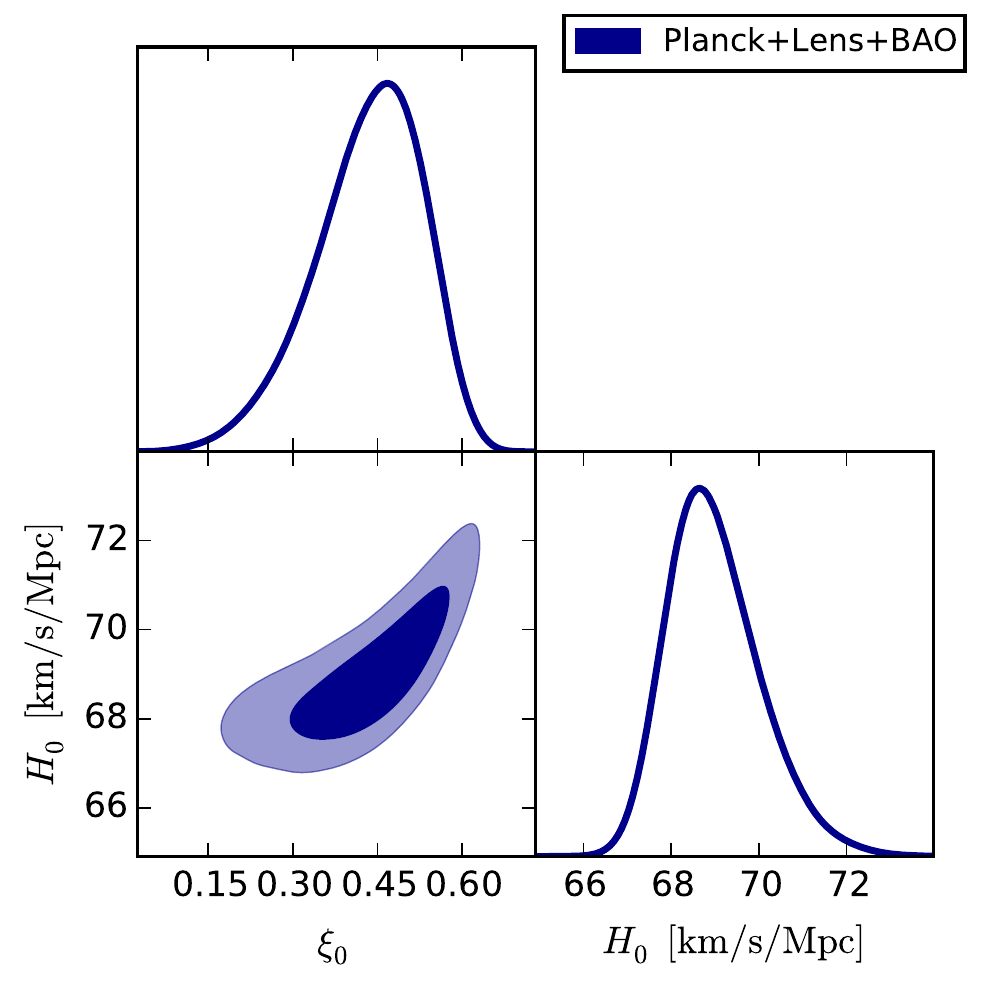}
\caption{Joint cosmological constraints on the present-day Hubble rate $H_0$ and the value of the dark photon to CMB photon temperature ratio $\xi_0$. The shaded blue regions show the $68\%$ and $95\%$ confidence regions using Planck 2015 CMB temperature and lensing data \cite{Ade:2015xua} as well as baryon acoustic oscillation (BAO) measurements from the 6dF galaxy survey \cite{2011MNRAS.416.3017B}, the Sloan Digital Sky Survey (SDSS) Main Galaxy Sample \cite{2015MNRAS.449..835R}, and the Baryon Oscillation Spectroscopic Survey (BOSS) DR11 \cite{2014MNRAS.441...24A}. Posterior distributions for $\xi_0$ and $H_0$ marginalized over all other cosmological and foreground parameters are shown along the diagonal. The methodology used to obtain this posterior distribution is identical to that used in ref.~\cite{Cyr-Racine:2013fsa}. We note that the Markov chains illustrated here were run with a flat prior on $\Delta N_{\rm eff}$, which is equivalent to putting a power-law prior on $\xi_0$.}
\label{fig:xi_vs_H0}
\end{center}
\end{figure}

The presence of a thermal bath of dark photons in our model adds an
extra contribution to the radiation energy density of the Universe. As
it is well known from standard cosmological parameter analyses (see
e.g.~\cite{Ade:2015xua}), adding extra relativistic species (commonly
parametrized by $\Delta N_{\rm eff}$) allows for a larger value of the
present-day Hubble parameter $H_0$ than that inferred using the
6-parameter $\Lambda$CDM model. While the properties of the dark
photons are different than the free-streaming relativistic species
parametrized by $\Delta N_{\rm eff}$ due their interactions with dark
matter \cite{Hou:2011ec,Cyr-Racine:2013fsa,2016JCAP...01..007B},
larger values of the Hubble parameters naturally occur in our model.
This is illustrated in figure~\ref{fig:xi_vs_H0} where we show the
$68\%$ and $95\%$ confidence regions for the $\xi_0-H_0$ parameter
space using Planck 2015 CMB temperature and lensing data
\cite{Ade:2015xua} as well as baryon acoustic oscillations
measurements (see figure caption for details). There we observe that
for the case where the dark and SM sectors are in thermal equilibrium
above the weak scale (corresponding to $\xi_0 \simeq 0.55$), values as
high as $H_0\simeq 72$ km/s/Mpc are within the $95\%$ confidence
region. Our model can therefore help reconciling CMB-based
measurements of the Hubble parameter with those from local probes
\cite{2016ApJ...826...56R}.

%

\section{Late-time astrophysics}
\label{sec:late-time-astro}
One of the key features of our model with a light charged $C$ particle is that it gives rise to dissipative dynamics, potentially leading to interesting astrophysical phenomena such as dark matter halo contraction and the formation of a dark matter disk~\cite{Fan:2013yva,Fan:2013tia}. In this section we review the physical mechanisms underpinning the dissipative dynamics of dark matter halos and identify the parameter space where it could occur. Compared to the original DDDM model, the fact that all dark matter halo particles are charged under the long-range force, and hence can all potentially lose energy through dissipation, is the main new feature of our model.
\subsection{Dissipative dynamics}
\label{sec:dissipative-dynamics}
A careful characterization of nonlinear structure formation in our model would require detailed numerical simulations taking into account the collisional and radiative processes that allow the different dark matter particles to lose, exchange, or gain energy and momentum. Similar to the case of hydrogen gas, the main mechanisms leading to energy lost among dark matter particles are Bremsstrahlung, inverse Compton scattering, collisional excitation and ionization, recombination and molecular cooling \cite{1996ApJS..105...19K}. Since the rate of collisional processes such as ionization and excitation are difficult to compute (see e.g.~refs.~\cite{1962PhRv..126..147B,Rosenberg:2017qia}), we present here a simplified approach and estimate the importance of dissipation within dark matter halos using Bremsstrahlung and inverse Compton scattering only. We expect these radiative processes to dominate when $T_C \gg B_{(XC)}$ or at early enough times when the energy density in dark photons is not entirely negligible. For $T_C\sim B_{(XC)}$, collisional processes including recombination cooling can become more important than radiative ones \cite{1981MNRAS.197..553B,1992ApJS...78..341C}. Since we are neglecting these former processes, our estimate should be taken as a lower bound on the possible amount of dissipation that can take place within a halo.

For the purpose of this calculation, we assume a dark matter halo with virial mass $M_{\rm vir} = 10^{12} M_{\odot}$ and virial radius $R_{\rm vir} = 100\ {\rm kpc}$. This implies an average density of $\bar{\rho}_{\rm DM}\simeq9\times10^{-3}\;\rm GeV/cm^3$. We perform our calculation at a benchmark redshift of $z=2$. We note that these choices are quite conservative since increasing the halo density or the benchmark redshift would both make dissipation more efficient. As dark matter and baryons fall in to form the halo, the dark plasma is shock heated to the virial temperature of the halo, which is approximately given by
\begin{align}
  T_{\rm vir}
  &=
  \frac{G_{\rm N} M_{\rm vir} \mu\, m_X}{5 R_{\rm vir}}
  \simeq \frac{9 \times 10^{2}}{1+f_{(XC)}} 
  \left(\frac{M_{\rm vir}}{10^{12} M_\odot}\right)
  \left(\frac{100 \ \rm{kpc}}{R_{\rm vir}}\right)
  \left(\frac {m_X}{\rm 10\ TeV}\right)\ {\rm keV} ,
\end{align}
where $\mu \equiv \rho_{\rm DM}/(2n_X m_X) \approx (1+f_{(XC)})^{-1}$ is the mean molecular weight of the dark plasma. Whenever $T_{\rm vir} > B_{(XC)}$, the $C$ particles are reionized. The free $C$ particles can then scatter off the ambient $X,\bar{X}$ particles, or off the dark photons, hence losing energy. The volumetric energy loss rate through Bremsstrahlung emission is given by \cite{Rybicki:847173}
\begin{align}
\Pi_{\rm Brem} &= \frac{16 \alpha_D^3\sqrt{2\pi T_C}}{(3 m_C)^{3/2} } n^{\rm free}_C (n^{\rm free}_X + n_{\bar{X}}),
\end{align}
while the equivalent rate for inverse Compton cooling is \cite{Seager:1999km}
\begin{align}
\Pi_{\rm Compt} &= \frac{64 \pi^3 \alpha_D^2 T_D^4}{135 m_C^3}n^{\rm free}_C T_C. 
\end{align}
Using these rates, we can define a typical cooling timescale $t_{{\rm cool,}C}$ for the free $C$ particles to shed an $\mathcal{O}(1)$ fraction of their kinetic energy through radiative processes
\begin{align}\label{eq:t_cool}
t_{{\rm cool,}C} &\equiv \left(\frac{2(\Pi_{\rm Brem} + \Pi_{\rm Compt})}{3 n_C^{\rm free} T_C}\right)^{-1}.
\end{align}
For this radiative cooling of the $C$ particles to have a significant impact on the structure of a dark matter halo, $t_{{\rm cool,}C}$ needs, at a minimum, to be  shorter than the age of the Universe $t_0$. We note that having $t_{{\rm cool,}C} < t_0$, with $t_{{\rm cool,}C}$ as defined in eq.~\eqref{eq:t_cool}, is a sufficient but not necessary condition for dissipation to play an important role in setting the internal structure of a dark matter halo since collisional processes could speed up the cooling of the $C$ particles. Equation \eqref{eq:t_cool} should thus be taken as an upper bound on the cooling timescale, given our choice of halo parameters. 

As the $C$ particles are dissipating their kinetic energy, they can scatter off $X$ and $\bar{X}$ particles in the halo. Since the net amount of kinetic energy in the $X$-$\bar{X}$ bath is larger than that of the $C$ particles (by a factor $\sim f_{(XC)}^{-1}T_X/T_C$), it is important to check that the heat transferred through Coulomb scattering from the $X$ and $\bar{X}$ to the $C$ particles does not significantly affect the latter's cooling. The volumetric heating rate of the $C$ particles from the $X$-$\bar{X}$ bath is \cite{1983HintonPlasma}
\begin{align}
\Pi_{\rm heat} & = \frac{4 \sqrt{\pi} \alpha_D^2 \ln \Lambda}{m_C m_X \left(\frac{2 T_C}{m_C} + \frac{2 T_X}{m_X} \right)^{3/2}} n_C^{\rm free} (n_X^{\rm free} + n_{\bar{X}}) T_X,
\end{align}
where $\ln\Lambda$ is the Coulomb logarithm as defined in eq.~\eqref{app:eq:coulomb_log}, and $T_X$ is the temperature of the $X$-$\bar{X}$ bath which is in general different than $T_C$. Defining the heating timescale as
\be
t_{{\rm heat},C}  \equiv \left(\frac{2\Pi_{\rm heat}}{3 n_C^{\rm free} T_C}\right)^{-1} ,
\ee 
we must have $t_{{\rm cool,}C} < t_{{\rm heat},C}$ in order for the $C$ particles to cool. By energy conservation, the $X$ and $\bar{X}$ particles must expend energy to heat the $C$ particles, leading to a net cooling of the heavy charged particle bath. The timescale associated with this $X$-$\bar{X}$ cooling, $t_{{\rm cool,}X}$, is in general much longer than $t_{{\rm heat,}C}$. Indeed,
\be
t_{{\rm cool,}X}  \equiv \left(\frac{2\Pi_{\rm heat}}{3 (n_X^{\rm free} + n_{\bar{X}}) T_X}\right)^{-1} = \frac{t_{{\rm heat},C} }{f_{(XC)}}\frac{T_X}{T_C},
\ee
where $1\lesssim T_X/T_C \lesssim T_{\rm vir}/B_{(XC)}$. For our parameter space of interest where $t_{{\rm cool,}C} < t_{{\rm heat},C}$, this implies that $t_{{\rm cool,}X} \gg t_{{\rm cool,}C}$ and the $X$-$\bar{X}$ bath is not significantly affected on the timescale at which the $C$ particles shed most of their kinetic energy. This is reassuring since the cooling of the $X$ and $\bar{X}$ particles in equilibrium with the $C$ particles would result in the collapse of the dark matter halo, which would be ruled out by observations. As another way to see that such equilibrium cooling does not happen in our scenario, notice that $t_{{\rm cool,}X} \sim t_{{\rm cool,}C}$ automatically implies that $t_{{\rm heat},C}\sim f_{(XC)} t_{{\rm cool,}C} \ll t_{{\rm cool,}C}$ if $f_{(XC)}\ll1$, meaning that little cooling happens since the heat from the $X$-$\bar{X}$ bath ends up warming up the $C$ particles faster than they can actually cool.

\begin{figure}[t]
\begin{center}
\includegraphics[width=0.49\textwidth]{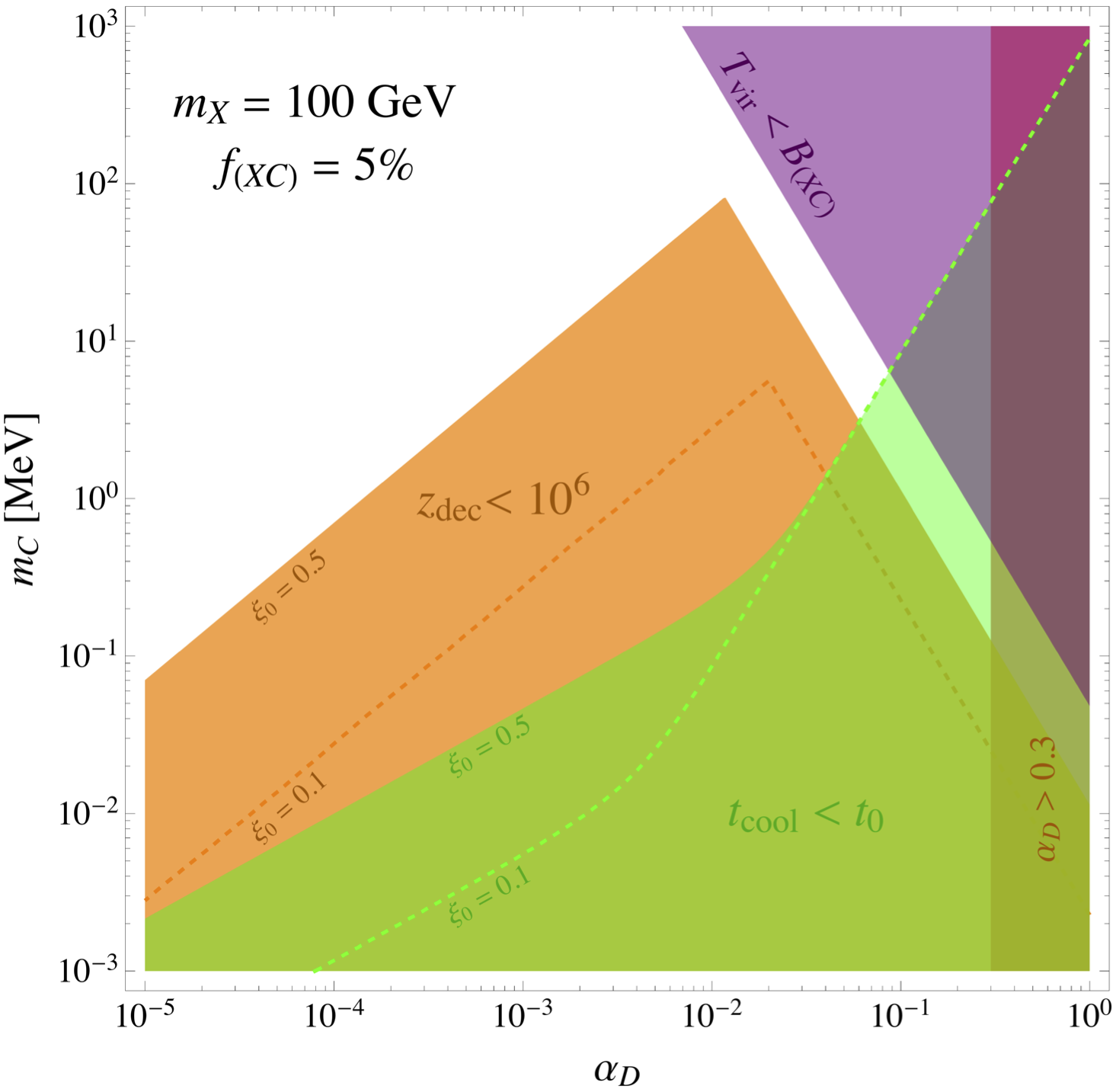}
\includegraphics[width=0.49\textwidth]{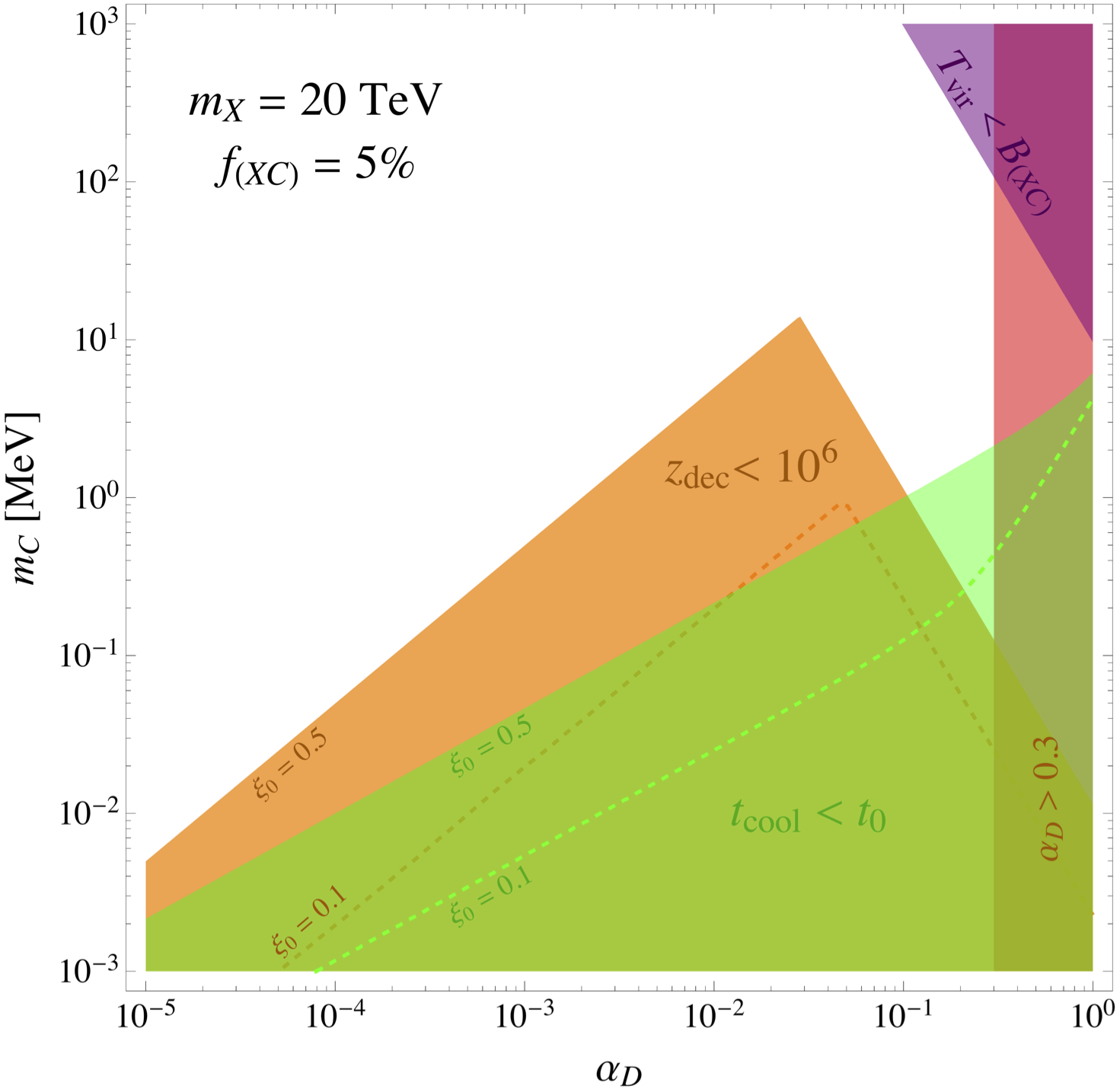}
\caption{Same as figure~\ref{fig:constraints} but fixing the $X$ mass to a constant value. The left panel has $m_X = 100$ GeV, while the right panel has $m_X = 20$ TeV. Both use $f_{(XC)} = 5\%$. }
\label{fig:constraints_fixed_mX}
\end{center}
\end{figure}

We illustrate in figs.~\ref{fig:constraints} and \ref{fig:constraints_fixed_mX} the parameter space (denoted by the light green regions) for which the typical timescale for the $C$ particles to lose most of their kinetic energy is less than the age of the Universe. We show the regions where $C$ cooling is possible for two different values of $\xi_0$ as labelled on the plots. It is immediately apparent that the parameter space where dissipative dynamics is possible has significant overlap with the parameter values ruled out by the abundance of small-scale structures. This illustrates the tension between structure formation and dissipation alluded to in section \ref{sec:model}: evading small-scale structure bounds pushes the $C$ particle mass toward higher values, while dissipation is most efficient for light $C$ particles. This tension is somewhat alleviated for large values of the $X$ mass, with values near $m_C\sim1$ MeV and $\alpha_D\sim0.15$ necessary to allow for significant dissipation while producing enough small-scale structures.

\subsection{Dark Matter Morphology}
Having established the region of parameter space where dissipative dynamics could play an important role, we now very briefly consider its impact on the dark matter distribution within bound halos. From section~\ref{sec:recombination} we know that for $\alpha_D \gtrsim 5\times 10^{-3}$ the protohalo is composed from relatively cold bound $(XC)$ states. For $\alpha_D \lesssim 5\times 10^{-3}$, the protohalo is made up by free $X$s and $C$s, with $C$s generally held at a common temperature. If the system virializes after infall into the forming galaxy (possibly due to shock heating), then the $X$s, $\bar{X}$s and $C$s thermalize at the same virial temperature $T_{\rm vir}$, making $C$s much faster than $X$s and $\bar{X}$s, namely $\sigma_C^2 \sim m_X/m_C \sigma_X^2$, where $\sigma_i^2 = \langle v_i^2 \rangle - \langle v_i \rangle^2$ is the velocity dispersion. However, while the $C$ can cool after the virialization, the temperature of $X$s stays nearly constant for most of our parameter space as mentioned in section~\ref{sec:dissipative-dynamics}. If the collision energy $T_{(XC)}$ in the center of mass of a typical $(XC)$ collision drops to $\sim B_{(XC)}$, the $C$s and $X$s recombine. To first order this happens when
\begin{align}
\alpha_D   \sim  (\sigma_C + \sigma_{\mathrm vir}).
\end{align}
As a result, if $\alpha_D \lesssim \sigma_{\mathrm vir} \sim 10^{-3}$, most $C$s stay ionized\footnote{If the $X$s move on average at velocities that are higher than typical velocities of $C$s in the bound state, $X$-$C$ collisions are unlikely to lead to bound states} no matter how much they cool, and the dark matter halo will be made out of unbound $X\bar{X}C$ plasma. This plasma can slowly cool through Compton and Bremsstrahlung processes potentially leading to a mildly flattened halo. However, as is apparent from figure~\ref{fig:constraints} it is unlikely to find a parameter point that allows for both efficient cooling and evades cosmological constraints for $\alpha_D < 10^{-3}$. Unless one extrapolates  to very small $\alpha_D$, the $X\bar{X}C$ plasma is unlikely. 

If, on the other hand, $\alpha_D \gtrsim 10^{-3}$, the $C$s will
rapidly recombine with $X$s as soon as $\sigma_C$ drops to about
$\alpha_D$, well before the Compton and Bremsstrahlung cooling lead to
a flattened halo. This halo will be formed from $(XC)$ atomic state
and unbound $X$, $\bar{X}$ plasma component. Although the $(XC)-X$
cross-section is not negligible forming a bound state reduces ability
of $C$s to act as a coolant and reduces the cooling rate of the halo.
As a result this scenario would not lead to a dark matter disk. As
mentioned in section~\ref{sec:dissipative-dynamics} there is parameter
space near $m_C\sim1$ MeV and $\alpha_D\sim0.15$ that realizes this
scenario.

However, this picture is woefully incomplete as we have evaluated the
cooling rates at specific benchmark density and fixed dark photon
temperature. Indeed, choosing a different fiducial halo mass, dark matter density, and collapse redshift would shift around the green regions shown in figs.~\ref{fig:constraints} and \ref{fig:constraints_fixed_mX}  where dissipative processes could occur. We have also completely neglected the effects of
collisional cooling/heating and recombination cooling (see ref.~\cite{Rosenberg:2017qia} for a recent computation of these rates). Taking all these effects into account, we generally expect dissipation to be efficient for only a limited range of halo masses. At the high-mass end, the large virial temperature of cluster-scale halos implies that the cooling timescale can become longer than the age of the Universe above a certain mass threshold, which depends on the choice of dark parameters. Small dwarf-size halos on the other hand can have a virial temperature that is below the $(XC)$ binding energy, hence suppressing dark matter dissipative processes. These expectations match the recent findings of ref.~\cite{Buckley:2017ttd} where they point out that the parameter space where dissipation occurs depends sensitively on the halo mass and density.  Moreover, we have not addressed the fact
that after virialization the velocity dispersion of the $C$ particles
is quite possibly relativistic and the $C$s could free stream out of
the galaxy, forming charge separation and subsequent large scale
electric fields. To what degree this becomes a significant effect is
unclear to us and in need of additional future analysis.     

The simple mechanism described above does not lead to formation of the
dark matter disk in 
the $XC$ model as was the case in DDDM \cite{Fan:2013tia,Fan:2013yva}. 
In order to form a disk, we would need a mechanism that effectively
couples only the asymmetric part of the $X$ population with the $C$s.

\section{Conclusions}
\label{sec:conclusions}

Dark matter can be charged under a new long range force, consistent with
everything observed to date. 
Cosmological constraints on darkly charged dark matter arise from the
dark matter coupling with the dark radiation, which suppresses the matter
power spectrum on small scales,
and from self-interactions within dark matter halos. Typically, darkly charged dark matter with a
weak-scale mass decouples from the dark radiation bath early enough to
 affect structure only at unobservably small scales, and can also evade self-interaction constraints \cite{Agrawal:2016quu}. However, in the
presence of a light particle charged under this new force (like our
$C$ particle here), dark matter would in general remain coupled to the
dark radiation bath through the epoch at which perturbations
corresponding to the smallest known dwarf galaxies enter the causal
horizon, or even through the time of CMB last scattering, hence leading to much more stringent constraints on these latter models. One possibility \cite{Fan:2013yva,Fan:2013tia} to evade these small-scale structure and CMB bounds \cite{Cyr-Racine:2013fsa} is to postulate that the darkly-charged dark matter makes up only a modest fraction of the overall dark matter density, with the rest being made of noninteracting cold dark matter. 

In this paper we considered the perhaps more economical possibility
that the rest of the dark matter is also made up of
the so-called heavy charged dark matter component $X$. 
Even though the $X$ particles scattering with dark radiation is
suppressed, they do have a large scattering cross section with
the light $C$ particles, which in turn couple strongly to photons.
Consequently, all of dark matter couples to dark radiation, potentially leading to
a strong suppression of the matter power spectrum on small scales.
In this model, dark matter must  kinetically decouple from the
dark radiation sufficiently early, $z_{\rm dec}\gtrsim10^6$, in order to be
consistent with small scale structure observations, e.g. dwarf
galaxies and Lyman-$\alpha$ measurements. This can happen if the
charge is so weak that decoupling happens sufficiently early or that
the binding energy of the $(XC)$ state is sufficiently high that the $C$ particles have recombined by $z=10^6$. Evading the structure formation constraint while allowing for dissipative dynamics is possible for $m_C\sim1$ MeV and large value of the coupling $\alpha_D\sim 0.15$. 
In this case, the
heavier dark matter, should it arise from a thermal freeze out, is
generally is required to be in the 10 TeV range. Our results are illustrated in figure~\ref{fig:constraints}, which is the
major result of this paper.

However, it is worth noting that even with early $C$ recombination or
small coupling, the effect of the new degrees of freedom, in the form
of the two additional dark photon polarizations, will remain. This can
have the effect of partially reconciling the apparent discrepancy in
measured Hubble parameters at late time and at the time of the CMB
imprinting (see figure~\ref{fig:xi_vs_H0}). It is also worth noting that for
relatively late decoupling of the dark and ordinary sectors at the
weak scale time, the dark radiation would have about half the
temperature of the SM bath, putting it at the border of being
discovered by better measurement \cite{Abazajian:2016yjj} of the number of relativistic degrees of
freedom ($N_{\rm eff}$). Future observations could reveal the suppression of the matter power
spectrum for models with $k_{\rm dec}\sim 10h/$Mpc. 
In addition, effects arising from
the long-range self-interactions between the unbound $X,\bar{X}$
particles in galactic haloes can lead to deviations from CDM
expectations. Finally, the presence of dark magnetic field in theories with a massless $U(1)$ gauge field could lead to new phenomena whose impact on structure formation is difficult to quantify (see e.g.~\cite{Ackerman:2008gi}).

In summary, darkly-charged dark matter is a very   interesting
possibility for our world. Given our limited knowledge about dark matter, it is
worthwhile investigating all possibilities, especially those that are
simple and testable. Darkly-charged dark matter on its own, and
especially coupled to a light dissipative component like the $C$
particle, could have a
significant impact on structure and cosmology which are well worth
exploring. Whether or not the WIMP paradigm proves correct, the TeV to
10 TeV range might turn out to be relevant not only to the Standard
Model, but to the dark sector too.

\begin{acknowledgments}
  We thank Adam Brown for suggesting our title. F.-Y. C.-R.~acknowledges the support of the National Aeronautical and Space Administration ATP grant NNX16AI12G at Harvard University. This work is supported by NSF grants PHY-0855591 and PHY-1216270. We would like to thank the Aspen Center for Physics, the Mainz Institute for Theoretical Physics, and David Rubenstein for hospitality during the completion of this work.
\end{acknowledgments}

\appendix
\section{Temperature evolution of the dark sector}\label{app:temp_evol}
\begin{figure}[t]
\begin{center}
\includegraphics[width=0.75\textwidth]{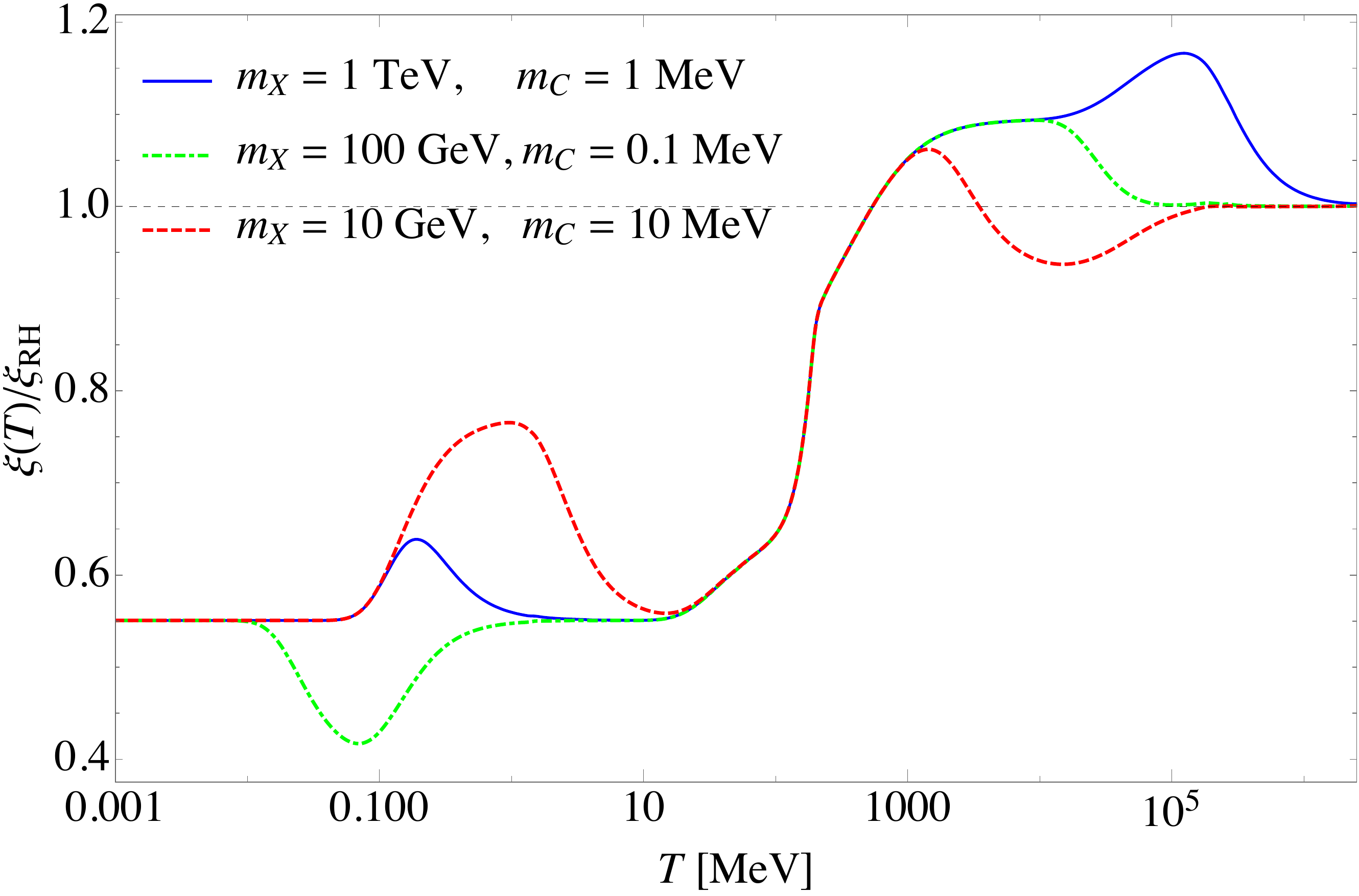}
\caption{Evolution of the ratio $\xi(T)$ of the dark sector to SM temperatures as given in eq.~\eqref{eq:xi_ratio} for a few representative $XC$ dark matter models. Here we take the values of $h_{\rm SM}(T)$ from ref.~\cite{Laine:2006cp}. We observe that depending on the $X$ and $C$ masses, we obtain a variety of behaviors. For instance, at temperatures $T<10$ MeV the $\xi(T)$ evolution depends strongly on whether the $C$ mass is below or above the electron mass. }
\label{Fig:xi_T}
\end{center}
\end{figure}

Given the value of the dark sector to SM temperature ratio at reheating $\xi_{\rm RH}$, it's value
at a later time is given by
 \begin{align}
   \label{eq:xi_ratio} 
   \xi(T) 
   &= 
   \left( \frac{h_{\rm SM}(T)}{h_{\rm SM}(T_{\rm RH})}
   \frac{h_D(T_{\rm RH})}{h_D(T)}
   \right)^{1/3}\xi_{\rm RH},
\end{align} 
where $T_{\rm RH}$ is the SM reheating temperature, $h_{\rm SM}$ is the effective number of degrees of freedom contributing to the entropy density of the SM, and $h_D$ is the similar quantity in the dark sector. We illustrate in figure~\ref{Fig:xi_T} examples of the evolution of $\xi$ for different choices of the $X$ and $C$ masses. For the models shown, the different behavior at temperatures above $1$ GeV is due to $X\bar{X}$ annihilation, while the evolution at temperatures less than $20$ MeV is caused by $C\bar{C}$ annihilation. We observe that even if both the SM and dark sector reheat to the same temperature $(\xi_{\rm RH} = 1)$, the  $X\bar{X}$ annihilation can briefly heat the dark sector above the SM temperature. However, the large entropy dump in the SM sector at the QCD phase transition generally results in a temperature ratio that is less than unity for the minimal dark matter model considered above. At late times, the temperature ratio takes the value $\xi(T_0)\equiv \xi_0 \simeq 0.55 \xi_{\rm RH}$, where $T_0=2.725$K is the temperature of the CMB today.
%

\section{Key definitions}\label{app:key_definition}
In section \ref{sec:relic-abundance}, we make use of the SM entropy density and the Hubble rate during radiation domination, which are respectively given by
\begin{align}
  s_{\rm SM}
  &=\frac{2\pi^2}{45} h_{\rm eff}(T) T^3,
  \\
  H &= \left[\frac{8\pi^3 }{90}\right]^\frac12 g_{\rm eff}^\frac12
  \frac{T^2}{ M_{\rm pl}},
\end{align}
where the Planck mass is $M_{\rm pl}=1.22\times10^{19}$ GeV. In deriving eq.~\eqref{eq:boltz}, we also used the time-temperature relation 
\begin{align}
  \frac{dt}{dx}
  &=
  \frac{dt}{dT}
  \frac{dT}{dx}
  =
  \frac{T^2}{m}
  \frac{1}{T H}
  \left(
  1+\frac{T}{3h_{\rm eff}}
  \frac{d h_{\rm eff} }{dT} 
  \right).
  \label{eq:timetemp}
\end{align}
The conformal time $\tau$ in a matter-radiation universe is related to the redshift as
\be\label{eq:app_tau_to_z}
\tau(z)  = \frac{2\sqrt{\Omega_{\rm rad}}}{H_0\Omega_{\rm m}}\left(\sqrt{1+\omega/(1+z)} -1\right),
\ee
where $\omega \equiv \Omega_{\rm m}/\Omega_{\rm rad}$. Here, $\Omega_{\rm m}$ and $\Omega_{\rm rad}$ is the total matter and radiation density of the universe in units of the critical density, respectively. 

\section{Evolution of dark matter density fluctuations}\label{app:DM_flucts}
We study here the evolution of the particle distribution function for the $X$, $\bar{X}$ and $C$ particles which we denote by $f_X(\pp{})$, $f_{\bar{X}}(\pp{})$ and $f_C(\pp{})$, respectively. Let us focus on the epoch after the chemical freeze-out of $X$ particles such that the processes $X\bar{X}\rightarrow\gamma_D \gamma_D$ and $X\bar{X}\rightarrow C\bar{C}$ are out of equilibrium and unimportant. The only relevant processes are then Coulomb interaction between the charged particles in the plasma and Thomson scattering between dark photons and the darkly charged particles.
\subsection{Boltzmann formalism}
 The set of coupled Boltzmann equations governing this system has the general form
\be
\left\{ \frac{df_i}{d\lambda} = \sum_{j} C_{ij\leftrightarrow ij}[f_i,f_j]\right\},
\ee
where $i,j \in \{X,\bar{X},C,\gamma_D\}$, and where $\lambda$ is an affine parameter describing the trajectory of the particle. The right-hand side of these equations are the collision terms defined with respect to $\lambda$. In the conformal Newtonian gauge, the space-time metric takes the form
\be\label{metric}
ds^2 = a^2(\tau)[-(1+2\psi) d\tau^2+(1-2\phi) d\vec{x}^2],
\ee
where $a$ is the cosmological scale factor, $\tau$ is the conformal time, and $\phi$ and $\psi$ are the two gravitational potentials. Note that we only include the \emph{scalar} fluctuations to the metric tensor since are the only ones that directly couple to density fluctuations at first order in perturbation theory. We can choose to define the affine parameter in terms of the four-momentum $P$ of the particle $P^\mu \equiv \frac{d x^\mu}{d\lambda}$, where $x^\mu=(\tau,\vec{x})$ is a four-vector parametrizing the world line of the particle.  We note that this implicitly sets the affine parameter to be the conformal time $\tau$ and selects a physically natural definition for the collision terms.  Using Eq.~(\ref{metric}), we can then write
 \be\label{eq:lambda}
 \frac{d}{d\lambda} = \frac{d\tau}{d\lambda}\frac{d}{d\tau}=P^0\frac{d}{d\tau}=\frac{E(1-\psi)}{a}\frac{d}{d\tau},
 \ee
 where we have used the dispersion relation $g_{\mu\nu} P^\mu P^\nu = -m^2$ and we have defined $E = \sqrt{p^2+m^2}$, $p= |\p|$, and $p^2 = g_{ij}P^iP^j$. We note that Eq.~(\ref{eq:lambda}) is valid to first-order in perturbation theory. The left-hand side of the above Boltzmann equations takes the form \cite{Dodelson-Cosmology-2003}
 \be\label{eq:LHS_Boltzmann_eq}
 \frac{df}{d\tau} =\frac{\pa f}{\pa \tau} + \frac{p}{E}\hat{p}^i\frac{\pa f}{\pa x^i}+p\frac{\pa f}{\pa p}\left[-\mathcal{H}+\frac{\pa \phi}{\pa \tau}-\frac{E}{p}\hat{p}^i\frac{\pa \psi}{\pa x^i}   \right],
\ee
 where in this work $\mathcal{H} = d\ln{a}/d\tau$ is the conformal Hubble expansion rate. It is useful to write down the bulk velocity $ \vec{v}_i$, number density $n_i$, and temperature $T_i$ of the $i$ particles ($i\in\{X,\bar{X},C,\gamma_D\}$) in terms of their distribution function:
\be\label{eq:DM_bulk_velocity}
 \vec{v}_i \equiv \frac{g_i}{n_i}\int\frac{d^3p}{(2\pi)^3}f_i(\pp{})\frac{\pp{}}{E},\qquad n_i \equiv g_i\int\frac{d^3p}{(2\pi)^3}f_i (\pp{})\qquad T_i \equiv \frac{g_i}{n_i}  \int\frac{d^3p}{(2\pi)^3}\frac{\pp{}^2}{3E}f_i(\pp{}),
\ee
where $g_i$ is the spin degeneracy factor for the $i$ particles.

To study the growth of dark matter fluctuations in this model, we need to evolve the equations for the number density perturbations $\de_i$ and the bulk velocities $\vec{v}_i$ of each charged constituents. Since Coulomb and Thomson scattering conserves the number of $X$, $\bar{X}$, $C$, and $\gamma_D$ particles, the collision integrals exactly vanish when integrated over all incoming momentum, that is,
\be
\int \frac{d^3p}{(2\pi)^3} C_{ij\leftrightarrow ij}(p) = 0,\quad \{i,j\}\in\{X,\bar{X},C,\gamma_D\}.
\ee
This implies that the density perturbations of the darkly charged constituents obey the standard evolution equation for cold dark matter \cite{Ma:1995ey,Cyr-Racine:2015ihg}
\be
\dot{\de}_i + \theta_i -3\dot{\phi}=0,
\ee
where an overhead dot denotes a derivative with respect to conformal time and where
\be\label{eq:def_delta_and_theta}
\de_i(\tau,\kk) \equiv \frac{n_i(\tau,\kk)}{\bar{n}_i(\tau)}-1, \qquad \theta_i(\tau,\kk) \equiv i\kk\cdot \vec{v}_i,
\ee
where $\bar{n}_i$ is the homogeneous and isotropic part of the $i$ number density. The vector $\kk$ is the Fourier wavenumber of the perturbation and it is understood that $\de_i$ and $\theta_i$ are evaluated in Fourier space. Using the definitions given in Eqs.~\eqref{eq:DM_bulk_velocity} and \eqref{eq:def_delta_and_theta}, the evolution equation for $\theta_i$ takes the general form \cite{Cyr-Racine:2015ihg}
\be\label{eq:DM_velocity}
\dot{\theta}_i-c_{i}^2k^2\de_i+\mathcal{H} \theta_i- k^2\psi =\frac{a(1+\psi)g_i}{n_i^{(0)}} \int \frac{d^3p}{(2\pi)^3}\frac{p(i\kk\cdot\hat{\bf p})}{E^2}  \sum_{j\neq i} C_{ij\leftrightarrow ij}(p),
\ee
where $c_i$ is the adiabatic sound speed of the $i$ dark constituents. For non-relativistic particles, we have $c_i\ll1$ and we can usually neglect this term unless $k$ is very large (corresponding to very small scales). Here, the collision integrals do not vanish in general since collisions will couple the bulk velocity of the different dark constituents. For the type of interactions that concern us here, it is possible to simplify the right-hand side of Eq.~\eqref{eq:DM_velocity} as \cite{Cyr-Racine:2015ihg}, 
\be
\dot{\theta}_i-c_{i}^2k^2\de_i+\mathcal{H} \theta_i- k^2\psi  = \sum_{j\neq i} \dot{\kappa}_{ij}(\theta_i - \theta_j), 
\ee
where the $\dot{\kappa}_{ij}$ are the opacities for the different processes. We provide in the next subsections explicit expressions for the relevant Thomson and Coulomb interactions between the constituents of our dark sector. 
 \subsection{Thomson Scattering}\label{app:sec_Thomson}
For Thomson scattering, the relevant opacities are computed in ref.~\cite{Cyr-Racine:2015ihg}. We however have to be careful to properly take into account the asymmetry and ionization fraction. The different Thomson opacities are then
 \be\label{eq:XC_gamma_scat}
 \dot{\kappa}_{C\gamma_D} = - \frac{4}{3}(\Omega_{\gamma_D}h^2) \frac{8\pi\alpha_D^2}{3 m_C^2}\frac{\tilde{\rho}_{\rm crit}}{m_C}R_{C}(z)(1+z)^3,
 \ee
  \be
 \dot{\kappa}_{\gamma_DC} = -(\Omega_{\rm DM}h^2) \frac{8\pi\alpha_D^2}{3 m_C^2}\frac{f_{(XC)}\tilde{\rho}_{\rm crit}}{m_C+m_X}R_{C}(z)(1+z)^2
 \ee

 \be
   \dot{\kappa}_{X\gamma_D} \simeq - \frac{4}{3}(\Omega_{\gamma_D}h^2) \frac{8\pi\alpha_D^2}{3 m_X^2}\frac{\tilde{\rho}_{\rm crit}}{m_X}\frac{1-f_{(XC)}+2f_{(XC)}R_{C}(z)}{1+f_{(XC)}}(1+z)^3,
 \ee
  \be
   \dot{\kappa}_{\gamma_DX} \simeq -(\Omega_{\rm DM}h^2) \frac{8\pi\alpha_D^2}{3 m_X^2}\frac{\tilde{\rho}_{\rm crit}}{m_X}\frac{1-f_{(XC)}+2f_{(XC)}R_{C}(z)}{2}(1+z)^2,
 \ee  
 \be\label{eq:gamma_barX_scat}
 \dot{\kappa}_{\bar{X}\gamma_D} = - \frac{4}{3}(\Omega_{\gamma_D}h^2) \frac{8\pi\alpha_D^2}{3 m_X^2}\frac{\tilde{\rho}_{\rm crit}}{m_X}(1+z)^3,
 \ee
 \be
 \dot{\kappa}_{\gamma_D\bar{X}} = -(\Omega_{\rm DM}h^2) \frac{8\pi\alpha_D^2}{3 m_X^2}\frac{\tilde{\rho}_{\rm crit}}{m_X}\frac{1-f_{(XC)}}{2}(1+z)^2.
 \ee
We note that the above expression are negative since we define our opacities as the actual derivative of the optical depth $\kappa$
 \be
 \kappa_{ij}(\tau) \equiv -\int_{\tau}^{\tau_0} d\tau' \dot{\kappa}_{ij}(\tau'),
 \ee
 where $\tau_0$ is the conformal time today.

 \subsection{Coulomb Scattering}\label{app:sec_Coulomb}
 Several approaches have been taken in computing the collision integrals for Coulomb scattering. An elegant starting point is to consider the scattering of a $X$ particle in a bath of $C$ particles as a Markov process, where the incoming state of the $X$ particle in any scattering event only depends on the previous scattering event. This immediately implies a collision term of the Fokker-Planck form \cite{Rosenbluth:1957zz}
 \be
 C[f(\pp{})] = - m \nabla_{\pp{}}\cdot \left[{\bf F} f(\pp{}) -\frac{1}{2}m \nabla_{\pp{}}\cdot \left( {\bf D} f(\pp{})\right)\right],
 \ee
 where ${\bf F}$ is the dynamical friction (or drift) vector and ${\bf D}$ is the diffusion tensor. Let us first consider the scattering of a $X$ particle on a bath of $C$ particles. The $C$ are light and scattering often off themselves and off dark photons, so their distribution function is close to a Maxwellian with a bulk velocity $\vec{v}_C$
 \be\label{eq:f_C}
 f_C(\pp{})\approx e^{(\mu_C-m_C)/T_C}e^{(-\frac{p^2}{2m_C}+\pp{}\cdot\vec{v}_C )/T_C},
 \ee 
 where $\mu_C$ is the $C$ chemical potential, and $T_C$ is the $C$ particle temperature, which is understood to have the form $T_C = T_{C,0}(\tau)(1+\de T_C(\xx,\tau))$. With this form, the Fokker-Planck term reduces to
 \be\label{eq:FK_coll_term}
 C[\fD(\pp{})] =  m_X\nabla_{\pp{}}\cdot \left[\gamma(\tau)\left((\pp{}-m_X \vec{v}_C)\fD(\pp{}) + m_X T_C  \nabla_{\pp{}} \fD(\pp{})\right)\right],
 \ee
 where the momentum transfer rate is \cite{Binder:2016pnr}
 \be
 \gamma(\tau) = \frac{1}{6 m_X T_C} \int \frac{d^3 \pp{2}}{(2\pi)^3} f_C(\pp{2})\int_{-4 \pp{2}^2}^0 dt\,(-t)\, \frac{d\sigma}{dt} v_{\rm rel}.
 \ee
 In the non-relativistic limit, we can write
 \be
 \frac{d\sigma}{dt} = \frac{1}{64 \pi m_X^2 p_2^2} |\bar{\mathcal{M}}|^2
 \ee
where
\be
\bar{|\mathcal{M}|^2} \equiv\frac{1}{g_X g_{C}}\sum_{\rm spins} |\mathcal{M}|^2,
\ee
where $g_X$ is the spin degeneracy of $X$. The matrix element for Coulomb scattering is well-known
\be\label{eq:coulomb_mat}
\frac{1}{g_X g_{C}}\sum_{\rm spins} |\mathcal{M}|^2 = \frac{128\pi^2 \alpha_D^2}{t^2}\left[ \left(\frac{s}{2}\right)^2 +  \left(\frac{u}{2}\right)^2-\frac{m_C^4-6 m_C^2 m_X^2 +m_X^4}{2}\right].
\ee
Averaging the Coulomb cross section over momentum transfer, we obtain
\be\label{eq:coulomb_ave}
\int_{-4 \pp{2}^2}^0 dt\,(-t)\, \frac{d\sigma}{dt} = \frac{12 \pi \alpha_D^2 m_C^2}{p_2^2} \ln\Lambda,
\ee
where $\ln\Lambda$ is the Coulomb logarithm. Using Eqs.~\eqref{eq:f_C} and \eqref{eq:coulomb_ave}, the momentum transfer rate $\gamma(\tau)$ is
\be
\gamma(\tau) = \frac{2\sqrt{2\pi} \alpha_D^2 \sqrt{\mu_{XC}} }{ m_X T_C^{3/2}} n_C \ln{\Lambda},
\ee
where 
\be
\mu_{XC} \equiv \frac{m_X m_C}{m_X+m_C}
\ee
is the reduced mass. Substituting Eq.~\eqref{eq:FK_coll_term} into Eq.~\eqref{eq:DM_velocity}, we obtain the opacities
\be\label{eq:XC_scat}
\dot{\kappa}_{XC} =  -a\frac{6\sqrt{2\pi} \alpha_D^2 \sqrt{\mu_{XC}} }{ m_X T_C^{3/2}} n_C^{\rm free} \ln{\Lambda},\qquad  \dot{\kappa}_{CX} =  -a\frac{6\sqrt{2\pi} \alpha_D^2 \sqrt{\mu_{XC}} }{ m_C T_C^{3/2}} n_X^{\rm free} \ln{\Lambda}.
\ee
For the Coulomb logarithm, we follow ref.~\cite{Rosenbluth:1957zz} and write
\be\label{app:eq:coulomb_log}
\Lambda = \frac{3}{4\pi}\left(\frac{T_C}{\alpha_D n_C}\right)^{3/2}n_C,
\ee
which is essentially the Debye screening length divided by the typical distance of closest approach. We note that if $T_C\propto (1+z)$ (as it is when C particles are thermally coupled to the dark photons), $\ln\Lambda$ is constant with values $\sim10-20$.


\bibliographystyle{JHEP.BST}
\bibliography{XC_references.bib}

\providecommand{\href}[2]{#2}\begingroup\raggedright\begin{thebibliography}{100}

\bibitem{1981ApJ...250..423D}
M.~{Davis}, M.~{Lecar}, C.~{Pryor} and E.~{Witten}, \emph{{The formation of
  galaxies from massive neutrinos}},
  \href{http://dx.doi.org/10.1086/159390}{\emph{\apj} {\bf 250} (Nov., 1981)
  423--431}.

\bibitem{Blumenthal:1982mv}
G.~R. Blumenthal, H.~Pagels and J.~R. Primack, \emph{{GALAXY FORMATION BY
  DISSIPATIONLESS PARTICLES HEAVIER THAN NEUTRINOS}},
  \href{http://dx.doi.org/10.1038/299037a0}{\emph{Nature} {\bf 299} (1982)
  37--38}.

\bibitem{Bond:1983hb}
J.~R. Bond and A.~S. Szalay, \emph{{The Collisionless Damping of Density
  Fluctuations in an Expanding Universe}},
  \href{http://dx.doi.org/10.1086/161460}{\emph{\apj} {\bf 274} (1983)
  443--468}.

\bibitem{Blumenthal:1984bp}
G.~R. Blumenthal, S.~Faber, J.~R. Primack and M.~J. Rees, \emph{{Formation of
  Galaxies and Large Scale Structure with Cold Dark Matter}},
  \href{http://dx.doi.org/10.1038/311517a0}{\emph{Nature} {\bf 311} (1984)
  517--525}.

\bibitem{Davis:1985rj}
M.~Davis, G.~Efstathiou, C.~S. Frenk and S.~D. White, \emph{{The Evolution of
  Large Scale Structure in a Universe Dominated by Cold Dark Matter}},
  \href{http://dx.doi.org/10.1086/163168}{\emph{\apj} {\bf 292} (1985)
  371--394}.

\bibitem{Goldberg:1986nk}
H.~Goldberg and L.~J. Hall, \emph{{A New Candidate for Dark Matter}},
  \href{http://dx.doi.org/10.1016/0370-2693(86)90731-8}{\emph{Phys. Lett.} {\bf
  B174} (1986) 151}.

\bibitem{HOLDOM198665}
B.~Holdom, \emph{Searching for ϵ charges and a new u(1)},
  \href{http://dx.doi.org/http://dx.doi.org/10.1016/0370-2693(86)90470-3}{\emph{Physics
  Letters B} {\bf 178} (1986) 65 -- 70}.

\bibitem{1992ApJ...398..407G}
B.-A. {Gradwohl} and J.~A. {Frieman}, \emph{{Dark matter, long-range forces,
  and large-scale structure}},
  \href{http://dx.doi.org/10.1086/171865}{\emph{\apj} {\bf 398} (Oct., 1992)
  407--424}.

\bibitem{1992ApJ...398...43C}
E.~D. {Carlson}, M.~E. {Machacek} and L.~J. {Hall}, \emph{{Self-interacting
  dark matter}}, \href{http://dx.doi.org/10.1086/171833}{\emph{\apj} {\bf 398}
  (Oct., 1992) 43--52}.

\bibitem{Foot:2003jt}
R.~Foot and R.~R. Volkas, \emph{{Was ordinary matter synthesized from mirror
  matter? An Attempt to explain why Omega(Baryon) approximately equal to 0.2
  Omega(Dark)}},
  \href{http://dx.doi.org/10.1103/PhysRevD.68.021304}{\emph{Phys. Rev.} {\bf
  D68} (2003) 021304}, [\href{http://arxiv.org/abs/hep-ph/0304261}{{\tt
  hep-ph/0304261}}].

\bibitem{Foot:2004pa}
R.~Foot, \emph{{Mirror matter-type dark matter}},
  \href{http://dx.doi.org/10.1142/S0218271804006449}{\emph{Int. J. Mod. Phys.}
  {\bf D13} (2004) 2161--2192},
  [\href{http://arxiv.org/abs/astro-ph/0407623}{{\tt astro-ph/0407623}}].

\bibitem{Foot:2004wz}
R.~Foot and R.~R. Volkas, \emph{{Spheroidal galactic halos and mirror dark
  matter}}, \href{http://dx.doi.org/10.1103/PhysRevD.70.123508}{\emph{Phys.
  Rev.} {\bf D70} (2004) 123508},
  [\href{http://arxiv.org/abs/astro-ph/0407522}{{\tt astro-ph/0407522}}].

\bibitem{Feng:2008mu}
J.~L. Feng, H.~Tu and H.-B. Yu, \emph{{Thermal Relics in Hidden Sectors}},
  \href{http://dx.doi.org/10.1088/1475-7516/2008/10/043}{\emph{JCAP} {\bf 0810}
  (2008) 043}, [\href{http://arxiv.org/abs/0808.2318}{{\tt 0808.2318}}].

\bibitem{Ackerman:2008gi}
L.~Ackerman, M.~R. Buckley, S.~M. Carroll and M.~Kamionkowski, \emph{{Dark
  Matter and Dark Radiation}},
  \href{http://dx.doi.org/10.1103/PhysRevD.79.023519}{\emph{Phys. Rev. D} {\bf
  79} (2009) 023519}, [\href{http://arxiv.org/abs/0810.5126}{{\tt 0810.5126}}].

\bibitem{Feng:2009mn}
J.~L. Feng, M.~Kaplinghat, H.~Tu and H.-B. Yu, \emph{{Hidden Charged Dark
  Matter}}, \href{http://dx.doi.org/10.1088/1475-7516/2009/07/004}{\emph{JCAP}
  {\bf 0907} (2009) 004}, [\href{http://arxiv.org/abs/0905.3039}{{\tt
  0905.3039}}].

\bibitem{ArkaniHamed:2008qn}
N.~Arkani-Hamed, D.~P. Finkbeiner, T.~R. Slatyer and N.~Weiner, \emph{{A Theory
  of Dark Matter}},
  \href{http://dx.doi.org/10.1103/PhysRevD.79.015014}{\emph{Phys. Rev. D} {\bf
  79} (2009) 015014}, [\href{http://arxiv.org/abs/0810.0713}{{\tt 0810.0713}}].

\bibitem{Kaplan:2009de}
D.~E. Kaplan, G.~Z. Krnjaic, K.~R. Rehermann and C.~M. Wells, \emph{{Atomic
  Dark Matter}},
  \href{http://dx.doi.org/10.1088/1475-7516/2010/05/021}{\emph{\jcap} {\bf
  1005} (2010) 021}, [\href{http://arxiv.org/abs/0909.0753}{{\tt 0909.0753}}].

\bibitem{2010PhRvD..81h3522B}
M.~R. {Buckley} and P.~J. {Fox}, \emph{{Dark matter self-interactions and light
  force carriers}},
  \href{http://dx.doi.org/10.1103/PhysRevD.81.083522}{\emph{Phys. Rev. D} {\bf
  81} (Apr., 2010) 083522}, [\href{http://arxiv.org/abs/0911.3898}{{\tt
  0911.3898}}].

\bibitem{Kaplan:2011yj}
D.~E. Kaplan, G.~Z. Krnjaic, K.~R. Rehermann and C.~M. Wells, \emph{{Dark
  Atoms: Asymmetry and Direct Detection}},
  \href{http://dx.doi.org/10.1088/1475-7516/2011/10/011}{\emph{\jcap} {\bf
  1110} (2011) 011}, [\href{http://arxiv.org/abs/1105.2073}{{\tt 1105.2073}}].

\bibitem{Behbahani:2010xa}
S.~R. Behbahani, M.~Jankowiak, T.~Rube and J.~G. Wacker, \emph{{Nearly
  Supersymmetric Dark Atoms}},
  \href{http://dx.doi.org/10.1155/2011/709492}{\emph{Adv. High Energy Phys.}
  {\bf 2011} (2011) 709492}, [\href{http://arxiv.org/abs/1009.3523}{{\tt
  1009.3523}}].

\bibitem{Das:2012aa}
S.~Das and K.~Sigurdson, \emph{Cosmological limits on hidden sector dark
  matter}, {\emph{Phys. Rev. D} {\bf 85} (2012) 063510},
  [\href{http://arxiv.org/abs/1012.4458}{{\tt 1012.4458}}].

\bibitem{Hooper:2012cw}
D.~Hooper, N.~Weiner and W.~Xue, \emph{{Dark Forces and Light Dark Matter}},
  \href{http://dx.doi.org/10.1103/PhysRevD.86.056009}{\emph{Phys. Rev. D} {\bf
  86} (2012) 056009}, [\href{http://arxiv.org/abs/1206.2929}{{\tt 1206.2929}}].

\bibitem{Aarssen:2012fx}
L.~G. van~den Aarssen, T.~Bringmann and C.~Pfrommer, \emph{{Is dark matter with
  long-range interactions a solution to all small-scale problems of $\Lambda$
  CDM cosmology?}},
  \href{http://dx.doi.org/10.1103/PhysRevLett.109.231301}{\emph{Phys. Rev.
  Lett.} {\bf 109} (2012) 231301}, [\href{http://arxiv.org/abs/1205.5809}{{\tt
  1205.5809}}].

\bibitem{Cline:2012is}
J.~M. Cline, Z.~Liu and W.~Xue, \emph{{Millicharged Atomic Dark Matter}},
  \href{http://dx.doi.org/10.1103/PhysRevD.85.101302}{\emph{Phys. Rev. D} {\bf
  85} (2012) 101302}, [\href{http://arxiv.org/abs/1201.4858}{{\tt 1201.4858}}].

\bibitem{Tulin:2013teo}
S.~Tulin, H.-B. Yu and K.~M. Zurek, \emph{{Beyond Collisionless Dark Matter:
  Particle Physics Dynamics for Dark Matter Halo Structure}},
  \href{http://dx.doi.org/10.1103/PhysRevD.87.115007}{\emph{Phys. Rev. D} {\bf
  87} (2013) 115007}, [\href{http://arxiv.org/abs/1302.3898}{{\tt 1302.3898}}].

\bibitem{Tulin:2012wi}
S.~Tulin, H.-B. Yu and K.~M. Zurek, \emph{{Resonant Dark Forces and Small Scale
  Structure}},
  \href{http://dx.doi.org/10.1103/PhysRevLett.110.111301}{\emph{Phys. Rev.
  Lett.} {\bf 110} (2013) 111301}, [\href{http://arxiv.org/abs/1210.0900}{{\tt
  1210.0900}}].

\bibitem{Baldi:2012ua}
M.~Baldi, \emph{{Structure formation in Multiple Dark Matter cosmologies with
  long-range scalar interactions}},
  \href{http://dx.doi.org/10.1093/mnras/sts169}{\emph{\mnras} {\bf 428} (2013)
  2074}, [\href{http://arxiv.org/abs/1206.2348}{{\tt 1206.2348}}].

\bibitem{Cyr-Racine:2013ab}
F.-Y. Cyr-Racine and K.~Sigurdson, \emph{The cosmology of atomic dark matter},
  {\emph{Phys. Rev. D} {\bf 87} (2013) 103515},
  [\href{http://arxiv.org/abs/1209.5752}{{\tt 1209.5752}}].

\bibitem{Cline:2013zca}
J.~M. Cline, Z.~Liu, G.~Moore and W.~Xue, \emph{{Composite strongly interacting
  dark matter}},
  \href{http://dx.doi.org/10.1103/PhysRevD.90.015023}{\emph{Phys. Rev. D} {\bf
  90} (2014) 015023}, [\href{http://arxiv.org/abs/1312.3325}{{\tt 1312.3325}}].

\bibitem{Baek:2013dwa}
S.~Baek, P.~Ko and W.-I. Park, \emph{{Hidden sector monopole, vector dark
  matter and dark radiation with Higgs portal}},
  \href{http://dx.doi.org/10.1088/1475-7516/2014/10/067}{\emph{JCAP} {\bf 1410}
  (2014) 067}, [\href{http://arxiv.org/abs/1311.1035}{{\tt 1311.1035}}].

\bibitem{Chu:2014lja}
X.~Chu and B.~Dasgupta, \emph{{Dark Radiation Alleviates Problems with Dark
  Matter Halos}},
  \href{http://dx.doi.org/10.1103/PhysRevLett.113.161301}{\emph{Phys. Rev.
  Lett.} {\bf 113} (2014) 161301}, [\href{http://arxiv.org/abs/1404.6127}{{\tt
  1404.6127}}].

\bibitem{Cline:2013pca}
J.~M. Cline, Z.~Liu, G.~Moore and W.~Xue, \emph{{Scattering properties of dark
  atoms and molecules}},
  \href{http://dx.doi.org/10.1103/PhysRevD.89.043514}{\emph{Phys. Rev. D} {\bf
  89} (2014) 043514}, [\href{http://arxiv.org/abs/1311.6468}{{\tt 1311.6468}}].

\bibitem{Bringmann:2013vra}
T.~Bringmann, J.~Hasenkamp and J.~Kersten, \emph{{Tight bonds between sterile
  neutrinos and dark matter}},
  \href{http://dx.doi.org/10.1088/1475-7516/2014/07/042}{\emph{JCAP} {\bf 1407}
  (2014) 042}, [\href{http://arxiv.org/abs/1312.4947}{{\tt 1312.4947}}].

\bibitem{Archidiacono:2014nda}
M.~Archidiacono, S.~Hannestad, R.~S. Hansen and T.~Tram, \emph{{Cosmology with
  self-interacting sterile neutrinos and dark matter - A pseudoscalar model}},
  \href{http://dx.doi.org/10.1103/PhysRevD.91.065021}{\emph{Phys. Rev. D} {\bf
  91} (2015) 065021}, [\href{http://arxiv.org/abs/1404.5915}{{\tt 1404.5915}}].

\bibitem{Foot:2014mia}
R.~Foot, \emph{{Mirror dark matter: Cosmology, galaxy structure and direct
  detection}}, \href{http://dx.doi.org/10.1142/S0217751X14300130}{\emph{Int. J.
  Mod. Phys.} {\bf A29} (2014) 1430013},
  [\href{http://arxiv.org/abs/1401.3965}{{\tt 1401.3965}}].

\bibitem{Foot:2014uba}
R.~Foot and S.~Vagnozzi, \emph{{Dissipative hidden sector dark matter}},
  \href{http://dx.doi.org/10.1103/PhysRevD.91.023512}{\emph{Phys. Rev.} {\bf
  D91} (2015) 023512}, [\href{http://arxiv.org/abs/1409.7174}{{\tt
  1409.7174}}].

\bibitem{Choquette:2015mca}
J.~Choquette and J.~M. Cline, \emph{{Minimal nonabelian model of atomic dark
  matter}},  \href{http://arxiv.org/abs/1509.05764}{{\tt 1509.05764}}.

\bibitem{Buen-Abad:2015ova}
M.~A. Buen-Abad, G.~Marques-Tavares and M.~Schmaltz, \emph{{Non-Abelian dark
  matter and dark radiation}},
  \href{http://dx.doi.org/10.1103/PhysRevD.92.023531}{\emph{Phys. Rev.} {\bf
  D92} (2015) 023531}, [\href{http://arxiv.org/abs/1505.03542}{{\tt
  1505.03542}}].

\bibitem{Ko:2016uft}
P.~Ko and Y.~Tang, \emph{{Light dark photon and fermionic dark radiation for
  the Hubble constant and the structure formation}},
  \href{http://dx.doi.org/10.1016/j.physletb.2016.10.001}{\emph{Phys. Lett.}
  {\bf B762} (2016) 462--466}, [\href{http://arxiv.org/abs/1608.01083}{{\tt
  1608.01083}}].

\bibitem{2016PhRvD..93l3527C}
F.-Y. {Cyr-Racine}, K.~{Sigurdson}, J.~{Zavala}, T.~{Bringmann},
  M.~{Vogelsberger} and C.~{Pfrommer}, \emph{{ETHOS -- an effective theory of
  structure formation: From dark particle physics to the matter distribution of
  the Universe}},
  \href{http://dx.doi.org/10.1103/PhysRevD.93.123527}{\emph{\prd} {\bf 93}
  (June, 2016) 123527}, [\href{http://arxiv.org/abs/1512.05344}{{\tt
  1512.05344}}].

\bibitem{Chacko:2016kgg}
Z.~Chacko, Y.~Cui, S.~Hong, T.~Okui and Y.~Tsai, \emph{{Partially Acoustic Dark
  Matter, Interacting Dark Radiation, and Large Scale Structure}},
  \href{http://dx.doi.org/10.1007/JHEP12(2016)108}{\emph{JHEP} {\bf 12} (2016)
  108}, [\href{http://arxiv.org/abs/1609.03569}{{\tt 1609.03569}}].

\bibitem{Ko:2016fcd}
P.~Ko and Y.~Tang, \emph{{Residual Non-Abelian Dark Matter and Dark
  Radiation}},
  \href{http://dx.doi.org/10.1016/j.physletb.2017.02.033}{\emph{Phys. Lett.}
  {\bf B768} (2017) 12--17}, [\href{http://arxiv.org/abs/1609.02307}{{\tt
  1609.02307}}].

\bibitem{Kamada:2016qjo}
A.~Kamada, K.~Kohri, T.~Takahashi and N.~Yoshida, \emph{{Effects of
  electrically charged dark matter on cosmic microwave background
  anisotropies}},
  \href{http://dx.doi.org/10.1103/PhysRevD.95.023502}{\emph{Phys. Rev.} {\bf
  D95} (2017) 023502}, [\href{http://arxiv.org/abs/1604.07926}{{\tt
  1604.07926}}].

\bibitem{Ko:2017uyb}
P.~Ko, N.~Nagata and Y.~Tang, \emph{{Hidden Charged Dark Matter and Chiral Dark
  Radiation}},  \href{http://arxiv.org/abs/1706.05605}{{\tt 1706.05605}}.

\bibitem{Foot:2002iy}
R.~Foot and S.~Mitra, \emph{{Ordinary atom mirror atom bound states: A New
  window on the mirror world}},
  \href{http://dx.doi.org/10.1103/PhysRevD.66.061301}{\emph{Phys. Rev.} {\bf
  D66} (2002) 061301}, [\href{http://arxiv.org/abs/hep-ph/0204256}{{\tt
  hep-ph/0204256}}].

\bibitem{Boehm:2000gq}
C.~Boehm, P.~Fayet and R.~Schaeffer, \emph{{Constraining dark matter candidates
  from structure formation}},
  \href{http://dx.doi.org/10.1016/S0370-2693(01)01060-7}{\emph{Phys. Lett.}
  {\bf B518} (2001) 8--14}, [\href{http://arxiv.org/abs/astro-ph/0012504}{{\tt
  astro-ph/0012504}}].

\bibitem{Boehm:2001hm}
C.~Boehm, A.~Riazuelo, S.~H. Hansen and R.~Schaeffer, \emph{{Interacting dark
  matter disguised as warm dark matter}},
  \href{http://dx.doi.org/10.1103/PhysRevD.66.083505}{\emph{Phys. Rev. D} {\bf
  66} (2002) 083505}, [\href{http://arxiv.org/abs/astro-ph/0112522}{{\tt
  astro-ph/0112522}}].

\bibitem{Boehm:2004th}
C.~Boehm and R.~Schaeffer, \emph{{Constraints on dark matter interactions from
  structure formation: Damping lengths}},
  \href{http://dx.doi.org/10.1051/0004-6361:20042238}{\emph{Astron. Astrophys.}
  {\bf 438} (2005) 419--442},
  [\href{http://arxiv.org/abs/astro-ph/0410591}{{\tt astro-ph/0410591}}].

\bibitem{Schewtschenko:2014fca}
J.~A. Schewtschenko, R.~J. Wilkinson, C.~M. Baugh, C.~Boehm and S.~Pascoli,
  \emph{{Dark matter-radiation interactions: the impact on dark matter
  haloes}}, \href{http://dx.doi.org/10.1093/mnras/stv431}{\emph{\mnras} {\bf
  449} (2015) 3587--3596}, [\href{http://arxiv.org/abs/1412.4905}{{\tt
  1412.4905}}].

\bibitem{Buckley:2014hja}
M.~R. Buckley, J.~Zavala, F.-Y. Cyr-Racine, K.~Sigurdson and M.~Vogelsberger,
  \emph{{Scattering, Damping, and Acoustic Oscillations: Simulating the
  Structure of Dark Matter Halos with Relativistic Force Carriers}},
  \href{http://dx.doi.org/10.1103/PhysRevD.90.043524}{\emph{Phys. Rev. D} {\bf
  90} (2014) 043524}, [\href{http://arxiv.org/abs/1405.2075}{{\tt 1405.2075}}].

\bibitem{Foot:2016wvj}
R.~Foot and S.~Vagnozzi, \emph{{Solving the small-scale structure puzzles with
  dissipative dark matter}},
  \href{http://dx.doi.org/10.1088/1475-7516/2016/07/013}{\emph{JCAP} {\bf 1607}
  (2016) 013}, [\href{http://arxiv.org/abs/1602.02467}{{\tt 1602.02467}}].

\bibitem{Foot:2011ve}
R.~Foot, \emph{{Mirror dark matter cosmology - predictions for $N_{eff} [CMB]$
  and $N_{eff} [BBN]$}},
  \href{http://dx.doi.org/10.1016/j.physletb.2012.04.023}{\emph{Phys. Lett.}
  {\bf B711} (2012) 238--243}, [\href{http://arxiv.org/abs/1111.6366}{{\tt
  1111.6366}}].

\bibitem{Cyr-Racine:2013fsa}
F.-Y. Cyr-Racine, R.~de~Putter, A.~Raccanelli and K.~Sigurdson,
  \emph{{Constraints on Large-Scale Dark Acoustic Oscillations from
  Cosmology}}, \href{http://dx.doi.org/10.1103/PhysRevD.89.063517}{\emph{Phys.
  Rev.} {\bf D89} (2014) 063517}, [\href{http://arxiv.org/abs/1310.3278}{{\tt
  1310.3278}}].

\bibitem{Foot:2013vna}
R.~Foot, \emph{{Galactic structure explained with dissipative mirror dark
  matter}}, \href{http://dx.doi.org/10.1103/PhysRevD.88.023520}{\emph{Phys.
  Rev.} {\bf D88} (2013) 023520}, [\href{http://arxiv.org/abs/1304.4717}{{\tt
  1304.4717}}].

\bibitem{Foot:2013lxa}
R.~Foot, \emph{{Tully-Fisher relation, galactic rotation curves and dissipative
  mirror dark matter}},
  \href{http://dx.doi.org/10.1088/1475-7516/2014/12/047}{\emph{JCAP} {\bf 1412}
  (2014) 047}, [\href{http://arxiv.org/abs/1307.1755}{{\tt 1307.1755}}].

\bibitem{Foot:2015sia}
R.~Foot, \emph{{Dissipative dark matter explains rotation curves}},
  \href{http://dx.doi.org/10.1103/PhysRevD.91.123543}{\emph{Phys. Rev.} {\bf
  D91} (2015) 123543}, [\href{http://arxiv.org/abs/1502.07817}{{\tt
  1502.07817}}].

\bibitem{Foot:2015mqa}
R.~Foot, \emph{{Dissipative dark matter and the rotation curves of dwarf
  galaxies}},
  \href{http://dx.doi.org/10.1088/1475-7516/2016/07/011}{\emph{JCAP} {\bf 1607}
  (2016) 011}, [\href{http://arxiv.org/abs/1506.01451}{{\tt 1506.01451}}].

\bibitem{Chashchina:2016wle}
O.~Chashchina, R.~Foot and Z.~Silagadze, \emph{{Radial acceleration relation
  and dissipative dark matter}},
  \href{http://dx.doi.org/10.1103/PhysRevD.95.023009}{\emph{Phys. Rev.} {\bf
  D95} (2017) 023009}, [\href{http://arxiv.org/abs/1611.02422}{{\tt
  1611.02422}}].

\bibitem{Foot:2017dgx}
R.~Foot, \emph{{Dissipative dark matter halos: The steady state solution}},
  \href{http://arxiv.org/abs/1707.02528}{{\tt 1707.02528}}.

\bibitem{Agrawal:2017pnb}
P.~Agrawal and L.~Randall, \emph{{Point Sources from Dissipative Dark Matter}},
   \href{http://arxiv.org/abs/1706.04195}{{\tt 1706.04195}}.

\bibitem{DAmico:2017lqj}
G.~D'Amico, P.~Panci, A.~Lupi, S.~Bovino and J.~Silk, \emph{{Massive Black
  Holes from Dissipative Dark Matter}},
  \href{http://arxiv.org/abs/1707.03419}{{\tt 1707.03419}}.

\bibitem{Fan:2013yva}
J.~Fan, A.~Katz, L.~Randall and M.~Reece, \emph{{Double-Disk Dark Matter}},
  \href{http://dx.doi.org/10.1016/j.dark.2013.07.001}{\emph{Phys. Dark Univ.}
  {\bf 2} (2013) 139--156}, [\href{http://arxiv.org/abs/1303.1521}{{\tt
  1303.1521}}].

\bibitem{Fan:2013tia}
J.~J. Fan, A.~Katz, L.~Randall and M.~Reece, \emph{{Dark-Disk Universe}},
  \href{http://dx.doi.org/10.1103/PhysRevLett.110.211302}{\emph{Phys. Rev.
  Lett.} {\bf 110} (2013) 211302}, [\href{http://arxiv.org/abs/1303.3271}{{\tt
  1303.3271}}].

\bibitem{Kahlhoefer:2013dca}
F.~Kahlhoefer, K.~Schmidt-Hoberg, M.~T. Frandsen and S.~Sarkar,
  \emph{{Colliding clusters and dark matter self-interactions}},
  \href{http://dx.doi.org/10.1093/mnras/stt2097}{\emph{Mon. Not. Roy. Astron.
  Soc.} {\bf 437} (2014) 2865--2881},
  [\href{http://arxiv.org/abs/1308.3419}{{\tt 1308.3419}}].

\bibitem{2016MNRAS.460.1399V}
M.~{Vogelsberger}, J.~{Zavala}, F.-Y. {Cyr-Racine}, C.~{Pfrommer},
  T.~{Bringmann} and K.~{Sigurdson}, \emph{{ETHOS - an effective theory of
  structure formation: dark matter physics as a possible explanation of the
  small-scale CDM problems}},
  \href{http://dx.doi.org/10.1093/mnras/stw1076}{\emph{\mnras} {\bf 460} (Aug.,
  2016) 1399--1416}, [\href{http://arxiv.org/abs/1512.05349}{{\tt
  1512.05349}}].

\bibitem{2016MNRAS.461..710D}
G.~A. {Dooley}, A.~H.~G. {Peter}, M.~{Vogelsberger}, J.~{Zavala} and
  A.~{Frebel}, \emph{{Enhanced tidal stripping of satellites in the galactic
  halo from dark matter self-interactions}},
  \href{http://dx.doi.org/10.1093/mnras/stw1309}{\emph{\mnras} {\bf 461}
  (Sept., 2016) 710--727}, [\href{http://arxiv.org/abs/1603.08919}{{\tt
  1603.08919}}].

\bibitem{Agrawal:2016quu}
P.~Agrawal, F.-Y. Cyr-Racine, L.~Randall and J.~Scholtz, \emph{{Make Dark
  Matter Charged Again}},
  \href{http://dx.doi.org/10.1088/1475-7516/2017/05/022}{\emph{JCAP} {\bf 1705}
  (2017) 022}, [\href{http://arxiv.org/abs/1610.04611}{{\tt 1610.04611}}].

\bibitem{Kaplan:2009ag}
D.~E. Kaplan, M.~A. Luty and K.~M. Zurek, \emph{{Asymmetric Dark Matter}},
  \href{http://dx.doi.org/10.1103/PhysRevD.79.115016}{\emph{Phys. Rev. D} {\bf
  79} (2009) 115016}, [\href{http://arxiv.org/abs/0901.4117}{{\tt 0901.4117}}].

\bibitem{Peebles:1968ja}
P.~Peebles, \emph{{Recombination of the Primeval Plasma}}, {\emph{Astrophys.J.}
  {\bf 153} (1968) 1}.

\bibitem{1968ZhETF..55..278Z}
Y.~B. {Zeldovich}, V.~G. {Kurt} and R.~A. {Syunyaev}, \emph{{Recombination of
  Hydrogen in the Hot Model of the Universe}}, {\emph{Zhurnal Eksperimental noi
  i Teoreticheskoi Fiziki} {\bf 55} (July, 1968) 278--286}.

\bibitem{2010MNRAS.406.1220W}
J.~{Wolf}, G.~D. {Martinez}, J.~S. {Bullock}, M.~{Kaplinghat}, M.~{Geha}, R.~R.
  {Mu{\~n}oz} et~al., \emph{{Accurate masses for dispersion-supported
  galaxies}},
  \href{http://dx.doi.org/10.1111/j.1365-2966.2010.16753.x}{\emph{\mnras} {\bf
  406} (Aug., 2010) 1220--1237}, [\href{http://arxiv.org/abs/0908.2995}{{\tt
  0908.2995}}].

\bibitem{2006PhRvL..97s1303S}
U.~{Seljak}, A.~{Makarov}, P.~{McDonald} and H.~{Trac}, \emph{{Can Sterile
  Neutrinos Be the Dark Matter?}},
  \href{http://dx.doi.org/10.1103/PhysRevLett.97.191303}{\emph{Physical Review
  Letters} {\bf 97} (Nov., 2006) 191303},
  [\href{http://arxiv.org/abs/astro-ph/0602430}{{\tt astro-ph/0602430}}].

\bibitem{2013PhRvD..88d3502V}
M.~{Viel}, G.~D. {Becker}, J.~S. {Bolton} and M.~G. {Haehnelt}, \emph{{Warm
  dark matter as a solution to the small scale crisis: New constraints from
  high redshift Lyman-{$\alpha$} forest data}},
  \href{http://dx.doi.org/10.1103/PhysRevD.88.043502}{\emph{\prd} {\bf 88}
  (Aug., 2013) 043502}, [\href{http://arxiv.org/abs/1306.2314}{{\tt
  1306.2314}}].

\bibitem{2015JCAP...11..011P}
N.~{Palanque-Delabrouille}, C.~{Y{\`e}che}, J.~{Baur}, C.~{Magneville},
  G.~{Rossi}, J.~{Lesgourgues} et~al., \emph{{Neutrino masses and cosmology
  with Lyman-alpha forest power spectrum}},
  \href{http://dx.doi.org/10.1088/1475-7516/2015/11/011}{\emph{\jcap} {\bf 11}
  (Nov., 2015) 011}, [\href{http://arxiv.org/abs/1506.05976}{{\tt
  1506.05976}}].

\bibitem{2016JCAP...08..012B}
J.~{Baur}, N.~{Palanque-Delabrouille}, C.~{Y{\`e}che}, C.~{Magneville} and
  M.~{Viel}, \emph{{Lyman-alpha forests cool warm dark matter}},
  \href{http://dx.doi.org/10.1088/1475-7516/2016/08/012}{\emph{\jcap} {\bf 8}
  (Aug., 2016) 012}, [\href{http://arxiv.org/abs/1512.01981}{{\tt
  1512.01981}}].

\bibitem{Ade:2015xua}
{\scshape Planck} collaboration, P.~A.~R. Ade et~al., \emph{{Planck 2015
  results. XIII. Cosmological parameters}},
  \href{http://dx.doi.org/10.1051/0004-6361/201525830}{\emph{Astron.
  Astrophys.} {\bf 594} (2016) A13},
  [\href{http://arxiv.org/abs/1502.01589}{{\tt 1502.01589}}].

\bibitem{2016JCAP...01..007B}
D.~{Baumann}, D.~{Green}, J.~{Meyers} and B.~{Wallisch}, \emph{{Phases of new
  physics in the CMB}},
  \href{http://dx.doi.org/10.1088/1475-7516/2016/01/007}{\emph{\jcap} {\bf 1}
  (Jan., 2016) 007}, [\href{http://arxiv.org/abs/1508.06342}{{\tt
  1508.06342}}].

\bibitem{2016ApJ...826...56R}
A.~G. {Riess}, L.~M. {Macri}, S.~L. {Hoffmann}, D.~{Scolnic}, S.~{Casertano},
  A.~V. {Filippenko} et~al., \emph{{A 2.4\% Determination of the Local Value of
  the Hubble Constant}},
  \href{http://dx.doi.org/10.3847/0004-637X/826/1/56}{\emph{\apj} {\bf 826}
  (July, 2016) 56}, [\href{http://arxiv.org/abs/1604.01424}{{\tt 1604.01424}}].

\bibitem{Ahn:2004xt}
K.-J. Ahn and P.~R. Shapiro, \emph{{Formation and evolution of the
  self-interacting dark matter halos}},
  \href{http://dx.doi.org/10.1111/j.1365-2966.2005.09492.x}{\emph{Mon. Not.
  Roy. Astron. Soc.} {\bf 363} (2005) 1092--1124},
  [\href{http://arxiv.org/abs/astro-ph/0412169}{{\tt astro-ph/0412169}}].

\bibitem{Moore:2000fp}
B.~Moore, S.~Gelato, A.~Jenkins, F.~R. Pearce and V.~Quilis, \emph{{Collisional
  versus collisionless dark matter}},
  \href{http://dx.doi.org/10.1086/312692}{\emph{Astrophys. J.} {\bf 535} (2000)
  L21--L24}, [\href{http://arxiv.org/abs/astro-ph/0002308}{{\tt
  astro-ph/0002308}}].

\bibitem{Randall:2007ph}
S.~W. Randall, M.~Markevitch, D.~Clowe, A.~H. Gonzalez and M.~Bradac,
  \emph{{Constraints on the Self-Interaction Cross-Section of Dark Matter from
  Numerical Simulations of the Merging Galaxy Cluster 1E 0657-56}},
  \href{http://dx.doi.org/10.1086/587859}{\emph{Astrophys. J.} {\bf 679} (2008)
  1173--1180}, [\href{http://arxiv.org/abs/0704.0261}{{\tt 0704.0261}}].

\bibitem{Graesser:2011wi}
M.~L. Graesser, I.~M. Shoemaker and L.~Vecchi, \emph{{Asymmetric WIMP dark
  matter}}, \href{http://dx.doi.org/10.1007/JHEP10(2011)110}{\emph{JHEP} {\bf
  10} (2011) 110}, [\href{http://arxiv.org/abs/1103.2771}{{\tt 1103.2771}}].

\bibitem{Gondolo:2004sc}
P.~Gondolo, J.~Edsjo, P.~Ullio, L.~Bergstrom, M.~Schelke and E.~A. Baltz,
  \emph{{DarkSUSY: Computing supersymmetric dark matter properties
  numerically}},
  \href{http://dx.doi.org/10.1088/1475-7516/2004/07/008}{\emph{JCAP} {\bf 0407}
  (2004) 008}, [\href{http://arxiv.org/abs/astro-ph/0406204}{{\tt
  astro-ph/0406204}}].

\bibitem{vonHarling:2014kha}
B.~von Harling and K.~Petraki, \emph{{Bound-state formation for thermal relic
  dark matter and unitarity}},
  \href{http://dx.doi.org/10.1088/1475-7516/2014/12/033}{\emph{JCAP} {\bf 1412}
  (2014) 033}, [\href{http://arxiv.org/abs/1407.7874}{{\tt 1407.7874}}].

\bibitem{1979rpa..book.....R}
G.~B. {Rybicki} and A.~P. {Lightman}, \emph{{Radiative processes in
  astrophysics}}.
\newblock {New York, Wiley-Interscience}, 1979.

\bibitem{Seager:1999km}
S.~Seager, D.~D. Sasselov and D.~Scott, \emph{{How exactly did the universe
  become neutral?}},
  \href{http://dx.doi.org/10.1086/313388}{\emph{Astrophys.J.Suppl.} {\bf 128}
  (2000) 407--430}, [\href{http://arxiv.org/abs/astro-ph/9912182}{{\tt
  astro-ph/9912182}}].

\bibitem{Dodelson-Cosmology-2003}
S.~Dodelson, \emph{Modern Cosmology}.
\newblock Academic Press. Academic Press, 2003.

\bibitem{1951ApJ...114..407S}
L.~{Spitzer}, Jr. and J.~L. {Greenstein}, \emph{{Continuous Emission from
  Planetary Nebulae}},
  \href{http://dx.doi.org/10.1086/145480}{\emph{Astrophys.J.} {\bf 114} (Nov.,
  1951) 407}.

\bibitem{Schewtschenko:2015rno}
J.~A. Schewtschenko, C.~M. Baugh, R.~J. Wilkinson, C.~B{\oe}hm, S.~Pascoli and
  T.~Sawala, \emph{{Dark matter--radiation interactions: the structure of Milky
  Way satellite galaxies}},
  \href{http://dx.doi.org/10.1093/mnras/stw1078}{\emph{Mon. Not. Roy. Astron.
  Soc.} {\bf 461} (2016) 2282--2287},
  [\href{http://arxiv.org/abs/1512.06774}{{\tt 1512.06774}}].

\bibitem{2011PhRvD..83d3506P}
E.~{Polisensky} and M.~{Ricotti}, \emph{{Constraints on the dark matter
  particle mass from the number of Milky Way satellites}},
  \href{http://dx.doi.org/10.1103/PhysRevD.83.043506}{\emph{\prd} {\bf 83}
  (Feb., 2011) 043506}, [\href{http://arxiv.org/abs/1004.1459}{{\tt
  1004.1459}}].

\bibitem{Cyr-Racine:2015ihg}
F.-Y. Cyr-Racine, K.~Sigurdson, J.~Zavala, T.~Bringmann, M.~Vogelsberger and
  C.~Pfrommer, \emph{{ETHOS - An Effective Theory of Structure Formation: From
  dark particle physics to the matter distribution of the Universe}},
  \href{http://arxiv.org/abs/1512.05344}{{\tt 1512.05344}}.

\bibitem{Ma:1995ey}
C.-P. Ma and E.~Bertschinger, \emph{{Cosmological perturbation theory in the
  synchronous and conformal Newtonian gauges}},
  \href{http://dx.doi.org/10.1086/176550}{\emph{\apj} {\bf 455} (1995) 7--25},
  [\href{http://arxiv.org/abs/astro-ph/9506072}{{\tt astro-ph/9506072}}].

\bibitem{Boehm:2003xr}
C.~Boehm, H.~Mathis, J.~Devriendt and J.~Silk, \emph{{Non-linear evolution of
  suppressed dark matter primordial power spectra}},
  \href{http://dx.doi.org/10.1111/j.1365-2966.2005.09032.x}{\emph{\mnras} {\bf
  360} (2005) 282--287}, [\href{http://arxiv.org/abs/astro-ph/0309652}{{\tt
  astro-ph/0309652}}].

\bibitem{Hou:2011ec}
Z.~Hou, R.~Keisler, L.~Knox, M.~Millea and C.~Reichardt, \emph{{How Massless
  Neutrinos Affect the Cosmic Microwave Background Damping Tail}},
  \href{http://dx.doi.org/10.1103/PhysRevD.87.083008}{\emph{Phys. Rev.} {\bf
  D87} (2013) 083008}, [\href{http://arxiv.org/abs/1104.2333}{{\tt
  1104.2333}}].

\bibitem{2011MNRAS.416.3017B}
F.~{Beutler}, C.~{Blake}, M.~{Colless}, D.~H. {Jones}, L.~{Staveley-Smith},
  L.~{Campbell} et~al., \emph{{The 6dF Galaxy Survey: baryon acoustic
  oscillations and the local Hubble constant}},
  \href{http://dx.doi.org/10.1111/j.1365-2966.2011.19250.x}{\emph{\mnras} {\bf
  416} (Oct., 2011) 3017--3032}, [\href{http://arxiv.org/abs/1106.3366}{{\tt
  1106.3366}}].

\bibitem{2015MNRAS.449..835R}
A.~J. {Ross}, L.~{Samushia}, C.~{Howlett}, W.~J. {Percival}, A.~{Burden} and
  M.~{Manera}, \emph{{The clustering of the SDSS DR7 main Galaxy sample - I. A
  4 per cent distance measure at z = 0.15}},
  \href{http://dx.doi.org/10.1093/mnras/stv154}{\emph{\mnras} {\bf 449} (May,
  2015) 835--847}, [\href{http://arxiv.org/abs/1409.3242}{{\tt 1409.3242}}].

\bibitem{2014MNRAS.441...24A}
L.~{Anderson}, {\'E}.~{Aubourg}, S.~{Bailey}, F.~{Beutler}, V.~{Bhardwaj},
  M.~{Blanton} et~al., \emph{{The clustering of galaxies in the SDSS-III Baryon
  Oscillation Spectroscopic Survey: baryon acoustic oscillations in the Data
  Releases 10 and 11 Galaxy samples}},
  \href{http://dx.doi.org/10.1093/mnras/stu523}{\emph{\mnras} {\bf 441} (June,
  2014) 24--62}, [\href{http://arxiv.org/abs/1312.4877}{{\tt 1312.4877}}].

\bibitem{1996ApJS..105...19K}
N.~{Katz}, D.~H. {Weinberg} and L.~{Hernquist}, \emph{{Cosmological Simulations
  with TreeSPH}}, \href{http://dx.doi.org/10.1086/192305}{\emph{\apjs} {\bf
  105} (July, 1996) 19}, [\href{http://arxiv.org/abs/astro-ph/9509107}{{\tt
  astro-ph/9509107}}].

\bibitem{1962PhRv..126..147B}
P.~G. {Burke} and H.~M. {Schey}, \emph{{Elastic Scattering of Low-Energy
  Electrons by Atomic Hydrogen}},
  \href{http://dx.doi.org/10.1103/PhysRev.126.147}{\emph{Physical Review} {\bf
  126} (Apr., 1962) 147--162}.

\bibitem{Rosenberg:2017qia}
E.~Rosenberg and J.~Fan, \emph{{Cooling in a Dissipative Dark Sector}},
  \href{http://arxiv.org/abs/1705.10341}{{\tt 1705.10341}}.

\bibitem{1981MNRAS.197..553B}
J.~H. {Black}, \emph{{The physical state of primordial intergalactic clouds}},
  \href{http://dx.doi.org/10.1093/mnras/197.3.553}{\emph{\mnras} {\bf 197}
  (Nov., 1981) 553--563}.

\bibitem{1992ApJS...78..341C}
R.~{Cen}, \emph{{A hydrodynamic approach to cosmology - Methodology}},
  \href{http://dx.doi.org/10.1086/191630}{\emph{\apjs} {\bf 78} (Feb., 1992)
  341--364}.

\bibitem{Rybicki:847173}
G.~B. Rybicki and A.~P. Lightman, \emph{{Radiative Processes in Astrophysics}}.
\newblock Wiley, New York, NY, 1985.

\bibitem{1983HintonPlasma}
F.~L. Hinton, \emph{Collisional Transport in Plasma}, vol.~1: Basic Plasma
  Physics of \emph{Handbook of Plasma Physics}.
\newblock North-Holland, New York, NY, 1983.

\bibitem{Buckley:2017ttd}
M.~R. Buckley and A.~DiFranzo, \emph{{Collapsed Dark Matter Structures}},
  \href{http://arxiv.org/abs/1707.03829}{{\tt 1707.03829}}.

\bibitem{Abazajian:2016yjj}
{\scshape CMB-S4} collaboration, K.~N. Abazajian et~al., \emph{{CMB-S4 Science
  Book, First Edition}},  \href{http://arxiv.org/abs/1610.02743}{{\tt
  1610.02743}}.

\bibitem{Laine:2006cp}
M.~Laine and Y.~Schroder, \emph{{Quark mass thresholds in QCD thermodynamics}},
  \href{http://dx.doi.org/10.1103/PhysRevD.73.085009}{\emph{Phys. Rev.} {\bf
  D73} (2006) 085009}, [\href{http://arxiv.org/abs/hep-ph/0603048}{{\tt
  hep-ph/0603048}}].

\bibitem{Rosenbluth:1957zz}
M.~N. Rosenbluth, W.~M. MacDonald and D.~L. Judd, \emph{{Fokker-Planck Equation
  for an Inverse-Square Force}},
  \href{http://dx.doi.org/10.1103/PhysRev.107.1}{\emph{Phys. Rev.} {\bf 107}
  (1957) 1--6}.

\bibitem{Binder:2016pnr}
T.~Binder, L.~Covi, A.~Kamada, H.~Murayama, T.~Takahashi and N.~Yoshida,
  \emph{{Matter Power Spectrum in Hidden Neutrino Interacting Dark Matter
  Models: A Closer Look at the Collision Term}},
  \href{http://dx.doi.org/10.1088/1475-7516/2016/11/043}{\emph{JCAP} {\bf 1611}
  (2016) 043}, [\href{http://arxiv.org/abs/1602.07624}{{\tt 1602.07624}}].

\end{thebibliography}\endgroup

\end{document}